\documentclass[12pt]{article}

\usepackage{latexsym}
\usepackage{amsmath,amssymb,exscale}
\usepackage{array,multicol}
\numberwithin{equation}{section}

\catcode`\@=11
\def\marginnote#1{}
\newcount\hour
\newcount\minute
\newtoks\amorpm
\hour=\time\divide\hour by60
\minute=\time{\multiply\hour by60 \global\advance\minute by-\hour}
\edef\standardtime{{\ifnum\hour<12 \global\amorpm={am}%
        \else\global\amorpm={pm}\advance\hour by-12 \fi
        \ifnum\hour=0 \hour=12 \fi
        \number\hour:\ifnum\minute<10 0\fi\number\minute\the\amorpm}}
\edef\militarytime{\number\hour:\ifnum\minute<10 0\fi\number\minute}
\def\draftlabel#1{{\@bsphack\if@filesw {\let\thepage\relax
   \xdef\@gtempa{\write\@auxout{\string
      \newlabel{#1}{{\@currentlabel}{\thepage}}}}}\@gtempa
   \if@nobreak \ifvmode\nobreak\fi\fi\fi\@esphack}
        \gdef\@eqnlabel{#1}}
\def\@eqnlabel{}
\def\@vacuum{}
\def\draftmarginnote#1{\marginpar{\raggedright\scriptsize\tt#1}}
\def\draft{\oddsidemargin -.5truein
        \def\@oddfoot{\sl preliminary draft \hfil
        \rm\thepage\hfil\sl\today\quad\militarytime}
        \let\@evenfoot\@oddfoot \overfullrule 3pt
        \let\label=\draftlabel
        \let\marginnote=\draftmarginnote
   \def\@eqnnum{(\theequation)\rlap{\kern\marginparsep\tt\@eqnlabel}%
\global\let\@eqnlabel\@vacuum}  }

\def\preprint{\twocolumn\sloppy\flushbottom\parindent 1em
        \leftmargini 2em\leftmarginv .5em\leftmarginvi .5em
        \oddsidemargin -.5in    \evensidemargin -.5in
        \columnsep 15mm \footheight 0pt
        \textwidth 250mmin      \topmargin  -.4in
        \headheight 12pt \topskip .4in
        \textheight 175mm
        \footskip 0pt
        \def\@oddhead{\thepage\hfil\addtocounter{page}{1}\thepage}
        \let\@evenhead\@oddhead \def\@oddfoot{} \def\@evenfoot{} }

\def\titlepage{\@restonecolfalse\if@twocolumn\@restonecoltrue\onecolumn
     \else \newpage \fi \thispagestyle{empty}\c@page\z@ 
        \def\thefootnote{\fnsymbol{footnote}} }

\def\endtitlepage{\if@restonecol\twocolumn \else  \fi
        \def\thefootnote{\arabic{footnote}}
        \setcounter{footnote}{0}}  %\c@footnote\z@ }

\catcode`@=12
\relax
\def\bea{\begin{array}}
\def\bem{\begin{displaymath}}
\def\beq{\begin{equation}}

\def\eea{\end{array}}
\def\eem{\end{displaymath}}
\def\eeq{\end{equation}}
\def\Im{\mathop{\rm Im}}

\def\ov{\overline}

\def\Re{\mathop{\rm Re}}
\def\s2w{\sin^2 \theta_W}

\relax
\def\dalpha{{\dot\alpha}}
\def\dbeta{{\dot\beta}}

\def\crbig{\\\noalign{\vspace {3mm}}}
\def\bigint{{\displaystyle\int}}

\def\Fint{{\bigint d^2\theta\,}}
\def\Fbarint{{\bigint d^2\ov\theta\,}}
\def\Dint{{\bigint d^2\theta d^2\ov\theta\,}}

\relax
\begin{document}

\topmargin-2.4cm
%\draft
%\preprint
%
%
%
%
\begin{titlepage}
\begin{flushright}
CERN-PH-TH/2008-068\\ 
CPHT-RR012.0408\\
April 2008
\end{flushright}
\vspace{1.3cm}

\begin{center}{\Large\bf
Nonlinear \boldmath{$N=2$} Supersymmetry, 
\\ 
\vspace{4mm}
Effective Actions and Moduli Stabilization}

\vspace{1.3cm}
{\large\bf I. Antoniadis$^{1,2}$,
%\footnote{On leave from CPHT (UMR CNRS 7644) Ecole Polytechnique, F-91128 Palaiseau}, 
J.-P. Derendinger$^3$ and T. Maillard$^2$}
\vspace{4mm}

\vskip .1in
$^1$ Department of Physics, CERN -- Theory Division\\ CH-1211 Geneva 23, Switzerland
\vskip .1in
$^2$ Centre de Physique Th\'eorique, UMR du CNRS 7644\\ Ecole Polytechnique, 91128 Palaiseau, France
\vskip .1in
$^3$ Physics Institute, Neuch\^atel University\\
Breguet 1, CH--2000 Neuch\^atel, Switzerland
\end{center}
\vspace{1.3cm}

\begin{center}
{\large\bf Abstract}
\end{center}
\begin{quote}
Nonlinear supersymmetry is used to compute the general form of the effective $D$-brane action
in type I string theory compactified to four dimensions in the presence of internal magnetic fields.
In particular, the scalar potential receives three contributions: (1) a nonlinear part of the D-auxiliary
component, associated to the Dirac-Born-Infeld action; (2) a Fayet-Iliopoulos (FI) D-term with a moduli-dependent coefficient; (3) a D-auxiliary independent (but moduli dependent) piece from the $D$-brane tension. Minimization of this potential leads to three general classes of vacua with moduli stabilization: (i) supersymmetric vacua allowing in general FI terms to be cancelled by non-trivial vacuum expectation values (VEV's) of charged scalar fields; (ii) anti-de Sitter vacua of broken supersymmetry in the presence of a non-critical dilaton potential that can be tuned at arbitrarily weak string coupling;
(iii) if the dilaton is fixed in a supersymmetric way by three-form fluxes and in the absence of charged scalar VEV's, one obtains non supersymmetric vacua with positive vacuum energy.

\end{quote}

\end{titlepage}
\setcounter{footnote}{0}
\setcounter{page}{0}
\setlength{\baselineskip}{.6cm}
\setlength{\parskip}{.2cm}
\newpage
%
% BODY
%

\section{Introduction}\label{secIntro}

$D$-branes break dynamically half of the bulk supersymmetries which are realized on their world-volume 
in a nonlinear way. This nonlinear supersymmetry can provide a powerful tool for computing the 
off-shell brane effective action~\cite{BG}. This is particularly important for moduli stabilization in the presence of 
internal magnetic fluxes. Indeed, for generic Calabi-Yau compactifications of type I string theory to 
four dimensions, magnetic fluxes generate a potential for closed string K\"ahler class moduli and 
play complementary role to three-form fluxes that generate a potential for the complex structure moduli 
and the dilaton~\cite{Giddings:2001yu, Kachru:2002he}, in order to stabilize all closed string moduli~\cite{KKLT} without introducing non-perturbative effects~\cite{Blumenhagen:2003vr, Antoniadis:2004pp, AKM2, Bianchi:2005yz}. In this work, we explore the tool of nonlinear ${\cal N}=2$ 
supersymmetry to determine the brane effective action and in particular the scalar potential.

We first rederive the Dirac-Born-Infeld (DBI) action from the ${\cal N}=2$ free Maxwell action by imposing 
the standard nonlinear constraint~\cite{RT, Rocek}, that relates its ${\cal N}=1$ chiral and vector multiplet components. 
Eliminating the former, one obtains the DBI action for the latter which is identified with the Goldstino 
multiplet of the second supersymmetry that becomes nonlinearly realized~\cite{BG}. Indeed, the corresponding 
transformations are modified by an additive constant piece. We can thus compute the off-shell scalar 
potential which is a nonlinear function of the ${\cal N}=1$ D-auxiliary field.

We then demonstrate that the ordinary ${\cal N}=1$ Fayet-Iliopoulos (FI) term is also invariant under the 
second nonlinear supersymmetry. Therefore, when added to the DBI potential, upon elimination of 
the D-auxiliary field, it yields an expression that depends nonlinearly on the coefficient of the FI-term 
which is a function of the closed string K\"ahler class moduli.

We finally show that the full potential contains an additive constant piece (from the point of view of 
global supersymmetry on the brane) arising from the brane tension, which we compute by taking 
into account the Ramond-Ramond (R-R) tadpole cancellation condition. Analyzing the resulting 
potential, including the dilaton factor, one finds three possible generic classes of minima:
\begin{enumerate}
\item A supersymmetric vacuum with K\"ahler class moduli stabilized by the vanishing condition for 
the coefficients of the FI-terms. An example of such complete stabilization is provided by the toroidal 
models of Refs.~\cite{Antoniadis:2004pp, AKM2, Antoniadis:2007jq}, in which case the complex structure 
moduli can also be 
stabilized by turning on magnetic fluxes on holomorphic two-cycles (which are absent in Calabi-Yau 
manifolds), leaving only the dilaton unfixed. However, these examples also show that sometimes the
supersymmetry conditions are incompatible with R-R tadpole cancellations, unless non-vanishing 
vacuum expectation values (VEV's) for charged scalar fields on the branes are turned on, as well. 
Obviously, these break partly the gauge symmetry on the branes.
\item Alternatively, in the absence of charged scalar VEV's, supersymmetry breaking vacua can be 
found. 
If the dilaton is already fixed in a supersymmetric way, for instance by three-form fluxes, these vacua 
are of de Sitter (dS) type with positive vacuum energy. 
\item Otherwise, by going off criticality in less than ten dimensions, one brings to the scalar potential 
an extra dilaton-dependent piece proportional to the central charge deficit, resulting to a 
supersymmetry breaking anti-de Sitter (AdS) vacuum with negative energy. This non-critical 
dilaton potential corresponds to a particular gauging of ${\cal N}=2$ effective supergravity associated 
to a shift isometry of the kinetic term of the universal (dilaton) hypermultiplet. The dilaton is then 
fixed at a value that can lead to an arbitrarily weak coupling, by making the central charge deficit 
infinitesimally small. A simple example is provided by replacing one free string coordinate with a 
conformal model from the minimal series.
\end{enumerate}

The organization of this paper is as follows. In Section~\ref{secN=2lin}, we review the main properties of 
${\cal N}=2$ linearly-realized supersymmetry. In particular, we describe the single tensor multiplet and 
the abelian vector multiplet in terms of two real ${\cal N}=1$ vector superfields. We then construct the 
${\cal N}=2$ superspace and derive the general form of an ${\cal N}=2$ supersymmetric action with 
arbitrary prepotential and FI terms of both electric and magnetic type, as well as the vector-tensor 
multiplet couplings. In Section~\ref{secN=2nonlin}, we describe the algorithm to obtain nonlinear 
${\cal N}=2$ supersymmetric actions. By imposing a nonlinear constraint, we derive the modification in 
the supersymmetry transformations, the DBI action and the corresponding off-shell scalar potential 
as a function of the D-auxiliary field. We also show that the ordinary FI term is invariant under 
the ${\cal N}=2$ nonlinear supersymmetry. In Section~\ref{secN=2sugra}, we discuss the ${\cal N}=2$ supergravity 
coupling and describe in particular the gauging that corresponds to adding a non-critical dilaton 
potential to be used in the following section. In Section~\ref{secpotential}, we apply our 
formalism to compute the scalar potential in four-dimensional type I string compactifications 
in the presence of internal magnetic fields. We then analyze this potential and find its 
supersymmetric and supersymmetry breaking minima. 
In addition, for self-consistency, there are three appendices. 
Appendix~\ref{A1} contains our conventions for ${\cal N}=1$ superspace, some useful identities 
involving super-covariant derivatives are listed in Appendix~\ref{A2} and, finally, Appendix~\ref{A3} 
presents the solution of the constraint for the nonlinear supersymmetry used in Section~\ref{secN=2nonlin},
following Ref. \cite{BG}.

% Section 2

\section{Linear \boldmath{${\cal N}=2$} supersymmetry} \label{secN=2lin}

The $D$-brane configurations we are interested in have linear ${\cal N}=1$ supersymmetry 
and a second nonlinearly-realized supersymmetry.
A convenient and simple formulation is then to use ${\cal N}=1$ superspace and construct 
linear ${\cal N}=2$ supersymmetry in this superspace. The nonlinear
realization is obtained by imposing constraints. 

In this context, there is a very simple way to realize linearly ${\cal N}=2$ supersymmetry on ${\cal N}=1$ superspace. Our conventions for ${\cal N}=1$ superspace are presented in Appendix~\ref{A1}. Start with two ${\cal N}=1$ superfields $V_i$, $i=1,2$. Under ${\cal N}=1$ 
supersymmetry,
$$ 
\delta V_i = (\epsilon Q + \ov\epsilon\ov Q ) \, V_i \, , \qquad\qquad (i=1,2),
$$
with spinor parameter $\epsilon$.
Then since 
$$
\{ Q_\alpha, \ov Q_\dalpha \} =  \{ D_\alpha, \ov D_\dalpha \}
= - 2i(\sigma^\mu)_{\alpha\dalpha}\partial_\mu 
$$
while all other anticommutators vanish, one can define a second supersymmetry
with spinor parameter $\eta$ by the transformations 
\beq
\label{linear*}
\delta^* V_1 = (a \, \eta D - \ov a \, \ov\eta\ov D) V_2 \, , \qquad\qquad
\delta^* V_2 = - (b \, \eta D - \ov b \, \ov\eta\ov D) V_1   \, ,
\eeq
with two complex numbers $a$ and $b$ such that $a\ov b = 1$. 
We will use the convention $a= -i/\sqrt2$ and $b=-\sqrt2i$. 
Notice that this procedure explicitly eliminates the $SU(2)$ covariance of the ${\cal N}=2$ 
supersymmetry algebra. 

Transformations (\ref{linear*}) provide an off-shell, linear realization of 
${\cal N}=2$ supersymmetry on ${\cal N}=1$ superfields $V_1$, $V_2$. In order to
describe an irreducible ${\cal N}=2$ multiplet, constraints compatible with 
both supersymmetries must be applied on $V_1$ and $V_2$.
For instance, if $V_1$ is chiral, then, since $\delta^* V_2 =  -b \eta DV_1$, 
$V_2$ cannot be chiral (or antichiral): this procedure cannot
be used to construct for instance the hypermultiplet 
(which does not admit an off-shell realization) using two chiral superfields. 
It can however be used to describe the single tensor multiplet and the 
abelian vector multiplet which are of direct interest for us. 
Since it will be needed in abelian gauge transformations, we begin with 
the single-tensor multiplet.

\subsection{The (single) tensor multiplet} \label{sectensor}

The ${\cal N}=2$ tensor multiplet~\cite{tensor1, LR, tensor2} describes an antisymmetric 
tensor, three real scalars and two Majorana spinors. These are the physical states 
of a chiral and a linear ${\cal N}=1$ superfields. 

Suppose that $V_1$ is a real linear superfield, $V_1=L$ with 
$DD \, L = \ov{DD}\, L =0$. Since 
$D_\alpha \,( \eta D \, L )  = \ov D_\dalpha \, ( \ov\eta\ov D \, L) = 0$,
the natural partner of $V_1 = L$ is then $V_2 = \phi+\ov\phi$, with
$\phi$ chiral and transformations
\beq
\label{*linear1}
\begin{array}{l}
\delta^* L = -\frac{i}{\sqrt2} \Bigl[ \eta D \, \phi + \ov\eta\ov D  \, \ov\phi \Bigr]
= -\frac{i}{\sqrt2} \Bigl[ \eta D + \ov\eta\ov D \Bigr] \, ( \phi+\ov\phi) , 
\crbig
\delta^*\phi = \sqrt2 i \, \ov\eta\ov D \, L, 
\qquad\qquad\qquad\qquad
\delta^*\ov\phi = \sqrt2 i \, \eta D \, L.
\end{array}
\eeq
A linear multiplet can be expressed in terms of a spinor chiral superfield
$\chi_\alpha$, $\ov D_\dalpha \chi_\alpha=0$\,\footnote{While $\chi_\alpha$
includes an antisymmetric tensor $b_{\mu\nu}$, $L$ includes its curl
$\partial_{[\mu} b_{\nu\rho]}$.}:
$$
L = D^\alpha\chi_\alpha - \ov D_\dalpha\ov\chi^\dalpha.
$$
It is defined up to the gauge transformation
$$
\chi_\alpha \quad\longrightarrow\quad \chi_\alpha + \ov{DD}D_\alpha U_\chi,
$$
for any real vector superfield $U_\chi$.
Instead of transformations (\ref{*linear1}), we may as well write
\beq
\label{*linear2}
\begin{array}{rclrcl}
\delta^*\chi_\alpha &=& -\frac{i}{\sqrt2} \, \phi \,  \eta_\alpha \,, \quad&\quad
\delta^*\ov\chi_\dalpha &=& \frac{i}{\sqrt2} \, \ov\phi \, \ov\eta_\dalpha \,,
\crbig
\delta^*\phi &=& 2\sqrt2 i\Bigl[\frac{1}{4}\, \ov{DD} \, \ov{\eta\chi} 
+ i \, \partial_\mu\chi\sigma^\mu\ov\eta \Bigr] ,
&
\delta^*\ov\phi &=& -2\sqrt2 i \Bigl[ \frac{1}{4} \, DD \, \eta\chi - i 
\, \eta\sigma^\mu\partial_\mu\ov\chi \Bigr] .
\end{array}
\eeq
Transformations (\ref{*linear1}) and (\ref{*linear2}) of the single-tensor 
${\cal N}=2$ supermultiplet were given by Lindstr\"om and Ro\v cek~\cite{LR}.

\subsection{The abelian vector multiplet} \label{secvector}

In ${\cal N}=1$ superspace, the ${\cal N}=2$ vector multiplet~\cite{vector} 
is commonly realized using the  gauge curvature chiral superfield 
$W_\alpha = - \frac{1}{4}\ov{DD}\, D_\alpha V$ and another chiral superfield $X$. 
The superfield $V$ is real (and dimensionless). In the Wess-Zumino gauge, it contains the
gaugino spinor $\lambda$, the real auxiliary scalar field $d$ and the gauge field:
$$
\begin{array}{rcl}
V &=& \theta\sigma^\mu\ov\theta \, A_\mu +i\,\theta\theta \, \ov\theta\ov\lambda 
-i\, \ov{\theta\theta} \, \theta\lambda + \frac{1}{2}\theta\theta\ov{\theta\theta} \, d \, , 
\crbig
W_\alpha &=& -i\lambda_\alpha + \theta_\alpha \, d + \frac{i}{2} 
(\theta\sigma^\mu\ov\sigma^\nu)_\alpha \, F_{\mu\nu} + \ldots, \qquad\qquad
F_{\mu\nu} = \partial_\mu A_\nu - \partial_\nu A_\mu \, .
\end{array}
$$
The superfield Bianchi identity is $D^\alpha W_\alpha = \ov D_\dalpha \ov W^\dalpha$.
The chiral $X$ has (mass) dimension one and since we consider the abelian theory, 
it is gauge invariant. Its components are the second gaugino $\psi_\alpha$, the complex 
scalar $x$ and a complex auxiliary scalar $f$: $X = x +\sqrt2\theta\psi - \theta\theta f$. 
The physical fields of the $N=2$ vector multiplets are then $(A_\mu, \psi, \lambda, x)$.
Under ${\cal N}=1$ supersymmetry, 
$$
\delta x = \sqrt 2 \, \epsilon\psi \, , \qquad\qquad 
\delta\psi_\alpha = -\sqrt2 i \, (\sigma^\mu\ov\epsilon)_\alpha \partial_\mu x 
+ \ldots
$$
(the auxiliary $f$ vanishes in the ${\cal N}=2$ super-Maxwell theory). We then also expect
$$
\delta^* x = \sqrt 2 \, \eta\lambda \, , \qquad\qquad \delta^*\lambda_\alpha = 
-\sqrt2 i \, (\sigma^\mu\ov\eta)_\alpha \partial_\mu x + \ldots
$$
for the second supersymmetry. 
This suggests to consider superfield variations of the form
$\delta^*X = \sqrt 2 \, i \, \eta^\alpha W_\alpha + \ldots$ and 
$\delta^* W_\alpha = - \sqrt 2 \, (\sigma^\mu\ov\eta)_\alpha \,\partial_\mu X + \ldots$
to realize the ${\cal N}=2$ superalgebra. 

To actually derive the second supersymmetry variations, 
suppose that $V_1$ and $V_2$ are two real vector superfields
and reduce their field content by imposing the following abelian 
gauge invariance:
\beq
\label{newgauge}
\begin{array}{rcll}
\delta_{gauge} \, V_1 &=& \Lambda_\ell \,, \qquad\qquad & DD\,\Lambda_\ell
= \ov{DD}\,\Lambda_\ell =0 \,,
\crbig
\delta_{gauge} \, V_2 &=& \Lambda_c + \ov \Lambda_c \,, \qquad\qquad &
\ov D_\dalpha\,\Lambda_c = D_\alpha\,\ov\Lambda_c =0 \, .
\end{array}
\eeq
While the gauge transformation of $V_2$ is as expected for a ${\cal N}=1$ abelian gauge superfield, $V_1$ transforms with a linear gauge parameter $\Lambda_\ell$.  
The second supersymmetry transformations
\beq
\label{vector*}
\delta^* V_1 = -\frac{i}{\sqrt2}\Bigl[ \eta D + \ov\eta\ov D \Bigr] V_2, \qquad\qquad
\delta^* V_2 = \sqrt2 i \Bigr[ \eta D + \ov\eta\ov D \Bigr] V_1
\eeq
are compatible with the gauge transformations since $\Lambda_c$ and 
$\Lambda_\ell$ form a tensor multiplet under $\delta^*$, with second 
supersymmetry transformations $\delta^*\Lambda_\ell$ and $\delta^*\Lambda_c$
as in eqs.~(\ref{*linear1}). Some useful identities involving covariant derivatives are
given in Appendix~\ref{A2}.
Define then the two gauge invariant superfields
\beq
\label{WXdef}
W_\alpha = -\frac{1}{4}\,\ov{DD}D_\alpha \, V_2 \, ,
\qquad \qquad
X = \frac{1}{2} \ov{DD}\, V_1 \, .
\eeq
Their variations under the second supersymmetry are
\beq
\label{*vector1}
\begin{array}{l}
\delta^*X = \sqrt 2 \, i \, \eta^\alpha W_\alpha ,
\qquad\qquad\qquad\qquad
\delta^*\ov X = \sqrt 2 \, i \, \ov\eta_\dalpha \ov W^\dalpha ,
\crbig
\delta^* W_\alpha =  \sqrt 2 \, i \left[ \frac{1}{4}\eta_\alpha \ov{DD}\,\ov X 
+ i (\sigma^\mu\ov\eta)_\alpha \, \partial_\mu X \right] ,
\crbig
\delta^* \ov W_\dalpha = \sqrt 2 \, i \, \left[ \frac{1}{4} \ov\eta_\dalpha {DD}\, X 
- i (\eta\sigma^\mu)_\dalpha \, \partial_\mu \ov X \right] .
\end{array}
\eeq
As a consequence of the definition of $W_\alpha$, they leave invariant 
the Bianchi identity $D^\alpha W_\alpha = \ov D_\dalpha\ov W^\dalpha$.

Then,
if $(W_\alpha,X)$ is the ${\cal N}=2$ supermultiplet of the abelian gauge curvature, 
$(V_1, V_2)$ with gauge invariance (\ref{newgauge}) gives the same
multiplet in terms of gauge fields (or gauge potentials). In the (generalized
to ${\cal N}=2$) Wess-Zumino gauge, the physical degrees of 
freedom of the supermultiplet are the gauge potential in the 
$\theta\sigma^\mu\ov\theta$ component of $V_2$, the two gauginos in the
$\theta\theta\ov\theta$ and $\ov{\theta\theta}\theta$ components of $V_1$ and 
$V_2$ and the complex scalar in the $\theta\theta$ and $\ov{\theta\theta}$
components of $V_1$. The $\theta\theta\ov{\theta\theta}$ components $d_1, d_2$
of $V_1$ and $V_2$ and the longitudinal vector in the $\theta\sigma^\mu\ov\theta$
component of $V_1$ are expected to be auxiliary. 

More precisely, 
choosing the Wess-Zumino gauge amounts to remove in the real vector superfield
$V_1$ a linear superfield $\Lambda_\ell$. In this gauge, $V_1$ reduces 
then to a Majorana fermion $\tilde\lambda$ (the second gaugino), a complex scalar 
$x$, the real (auxiliary) 
scalar $d_1$ and the longitudinal component of a vector field $\tilde v_\mu$:
$$
\left. V_1 \right|_{W.Z.} = -\frac{1}{2}\theta\theta \, \ov x - \frac{1}{2} \ov{\theta\theta} \, x
+ \theta\sigma^\mu\ov\theta \, \tilde v_\mu
- \frac{i}{\sqrt2}\theta\theta \ov\theta\ov{\tilde\lambda} 
- \frac{i}{\sqrt2}\ov{\theta\theta} \theta\tilde\lambda 
+  \frac{1}{2} \theta\theta \ov{\theta\theta} \, d_1,
$$
with residual gauge invariance $\delta\tilde v^\mu = 
\epsilon^{\mu\nu\rho\sigma}\partial_\nu\Lambda_{\rho\sigma}$
to eliminate the transverse part of $\tilde v_\mu$.
With relation $X=\frac{1}{2}\ov{DD}V_1$ (and in chiral variables),
$$
X = x + \sqrt2\, \theta\tilde\lambda - \theta\theta f_X, 
\qquad\qquad
f_X = d_1 + i \partial^\mu \tilde v_\mu \, .
$$
In a theory depending on $X$ only, as in the super-Maxwell theory, replacing
$\Im f_X$ by the field $\partial^\mu \tilde v_\mu$ has a single implication:
a linear term $cX$ ($c$ complex) in the superpotential, which potentially breaks
supersymmetry, reduces now to a term $\Re c \Re X$, with the same supersymmetry 
breaking pattern. In other words, replacing $\Im f_X$ by $\partial^\mu \tilde v_\mu$ is 
equivalent to a choice of phase of the term linear in $X$ in the superpotential. 

\subsection{The \boldmath{${\cal N}=2$} super-Maxwell theory} \label{secMaxwell}

The ${\cal N}=2$ super-Maxwell theory with Lagrangian
\beq
\label{N=2Maxwell}
\begin{array}{rcl}
{\cal L}_{Max.} &=& \Dint \ov XX + \frac{1}{4} \Fint WW + \frac{1}{4}\Fbarint \ov{WW}
\crbig
&=& \frac{1}{4} \Fint\left[ WW - \frac{1}{2} X \ov{DD}\ov X \right] + {\rm h.c.}
+ {\rm total\ derivative}
\end{array}
\eeq
is invariant under the second supersymmetry: from variations (\ref{*vector1}), one
obtains
\beq
\label{*susy1}
\delta^* \left[ WW - \frac{1}{2} X \ov{DD}\ov X \right] = - 2\sqrt2 \, \partial_\mu
(W\sigma^\mu\ov\eta \, X),
\eeq
a total derivative leading to an invariant action. 

With two vector superfields $V_1$ and $V_2$ and a chiral $X$, 
the following supplementary terms are ${\cal N}=2$ and gauge invariant:
\beq
\label{FI1}
{\cal L}_{F.I.} = \Dint [\xi_1V_1 + \xi_2V_2] + 
\zeta\Fint X + {\rm h.c.},
\eeq
with $\xi_1$ and $\xi_2$ real and $\zeta$ complex. 
Since the $\theta\theta$ component of $W_\alpha$ is a total derivative,
$$
\delta^* \Fint \zeta X = \sqrt2 i \,\zeta\, \Fint W\eta
$$
is a total derivative, but a more complicated superpotential in $X$ is
forbidden by the second supersymmetry. 
However, since $X=\frac{1}{2}\ov{DD}V_1$,
$$
\zeta\Fint X + {\rm h.c.} = -4(\Re \zeta)\Dint V_1 + \makebox{total\ derivative},
$$
which indicates that the imaginary part of $\zeta$ is irrelevant 
while $\Re\zeta$ and $\xi_1$ are redundant. 
The theory has then two Fayet-Iliopoulos terms
\beq
\label{FI2}
{\cal L}_{F.I.} = \Dint [\xi_1V_1 + \xi_2V_2]
= -\frac{1}{2} \xi_1\Re\Fint X + \xi_2 \Dint V_2 ,
\eeq
with two real arbitrary parameters.
In the Wess-Zumino gauge, 
\beq
\label{FI3}
{\cal L}_{F.I.} = \frac{1}{2} \left[ \xi_1d_1 + \xi_2d_2 \right]
\eeq
which, if added to an interacting theory with a non-trivial prepotential (see eq.~(\ref{LMax}) below),  generates a positive
scalar potential breaking both supersymmetries with the same order parameter
$\sqrt{\xi_1^2 + \xi_2^2}$. 

\subsection{\boldmath{${\cal N}=2$} superspace construction and 
prepotential}\label{secprepot}

For completeness, we relate the above derivation of the supermultiplet 
$(W_\alpha, X)$ with its familiar construction in ${\cal N}=2$ superspace~\cite{GSW}. Start
with a chiral ${\cal N}=2$ superfield in chiral coordinates $(y,\theta,\tilde\theta)$
and expand in $\tilde\theta$:
$$
\Phi (y,\theta,\tilde\theta) = X(y,\theta) + i\sqrt2 \, \tilde\theta W(y,\theta)
- \tilde\theta\tilde\theta \, F(y,\theta).
$$
We know from ${\cal N}=1$ superspace that
\beq
\label{*chiral}
\begin{array}{rcl}
\delta^* X &=& \sqrt 2 i \, \eta W , 
\crbig
\delta^* W_\alpha &=& \sqrt2 i \, [ F \eta_\alpha  
+ i(\sigma^\mu\ov\eta)_\alpha \partial_\mu X]  ,
\crbig
\delta^* F &=& \sqrt2  \, \partial_\mu W \sigma^\mu\ov\eta.
\end{array}
\eeq
Since
$$
\frac{1}{4}\int d^2\tilde\theta \, \Phi^2
= \frac{1}{4}\left[ WW - 2XF \right],
$$
the ${\cal N}=2$ super-Maxwell system (\ref{N=2Maxwell}) is recovered if we impose
\beq
\label{Fcond}
F = \frac{1}{4} \, \ov{DD}\ov X.
\eeq
This condition is actually compatible with the supersymmetry variations (\ref{*chiral})
and eq.~(\ref{Fcond}), when inserted into the first two eqs.~(\ref{*chiral}), leads again
to the transformations (\ref{*vector1}).

This derivation of the free ${\cal N}=2$ Maxwell theory easily 
generalizes to an interacting model using the holomorphic prepotential 
${\cal F}(\Phi)$:
\beq
\label{LMax}
\begin{array}{rcl}
{\cal L}_{\cal F} &=& \frac{m^2}{2}\bigint d^2\theta 
\bigint d^2\tilde\theta \, {\cal F}(\Phi/m) + {\rm h.c.}
\crbig
&=& \frac{1}{4}\Fint \left[ {\cal F}^{\prime\prime}(X/m) \, WW 
-\frac{m}{2} \, {\cal F}^\prime(X/m) \, \ov{DD}\ov X \right] + {\rm h.c.}
\crbig
&=& \frac{m}{2}\Dint \left[{\cal F}^\prime(X/m) \, \ov X 
+ \ov {\cal F}^\prime(\ov X/m) \, X  \right] 
+ \frac{1}{4}\Fint  {\cal F}^{\prime\prime}(X/m) \, WW  + {\rm h.c.},
\end{array}
\eeq
where $m$ is an arbitrary (real) mass scale\footnote{The scale $m$ 
disappears in the free, scale-invariant, case with quadratic prepotential, 
${\cal F} = X^2/2m^2$.}.
Invariance under the second supersymmetry, with variations (\ref{*vector1}), 
follows from 
\beq
\label{*susy2}
\delta^*  \left[ {\cal F}^{\prime\prime}(X/m)WW 
-\frac{m}{2}{\cal F}^\prime(X/m) \ov{DD}\ov X \right] 
= -2\sqrt2 \, \partial_\mu \Bigl[ {\cal F}^\prime(X/m) W\sigma^\mu\ov\eta  \Bigr].
\eeq
The Fayet-Iliopoulos terms (\ref{FI2}), which break spontaneously both supersymmetries, 
can be added to Lagrangian (\ref{LMax}). In addition, 
as demonstrated in Ref.~\cite{APT}, this combined theory admits 
a deformation in which one supersymmetry is nonlinearly realized (in 
the ``Goldstino mode") and supersymmetry partially breaks.\footnote{For a generalization to the non-abelian case see Ref.~\cite{Fujiwara:2004kc}.}

\subsection{Vector--tensor multiplet couplings} \label{secvectortensor}

To couple a tensor multiplet $(L,\phi)$ to the vector multiplet
$(V_1,V_2)$, we may postulate gauge variations
\beq
\label{gaugetensor}
\delta_{gauge} \, L = h \Lambda_\ell \,, \qquad\qquad
\delta_{gauge} \, \phi = h \Lambda_c ,
\eeq
where the real number $h$ plays the role of a charge. 
The gauge invariant combinations 
\beq
\label{hatVs}
\hat V_1 = L- hV_1\,, \qquad\qquad
\hat V_2 = \phi + \ov\phi - hV_2
\eeq
verify then
\beq
\label{delta*Vs}
\delta^*\hat V_1 =  -\frac{i}{\sqrt2} \, \Bigl[ \eta D + \ov\eta\ov D \Bigr] \,\hat V_2 \,,
\qquad\qquad
\delta^*\hat V_2 =  \sqrt2 i \, \Bigl[ \eta D + \ov\eta\ov D \Bigr] \,\hat V_1\,.
\eeq
Since
$$
\delta^*\biggl(\hat V_1 \pm \frac{i}{\sqrt2} \hat V_2 \biggr) =
\mp \, \Bigl[ \eta D + \ov\eta\ov D \Bigr] 
\biggl(\hat V_1 \pm \frac{i}{\sqrt2} \hat V_2\biggr) ,
$$
one immediately infers that
\beq
\label{coupling1}
\Dint \left[ {\cal H}(\hat V_1 + i \hat V_2 / \sqrt2) 
+{\cal H}^*(\hat V_1 - i \hat V_2 / \sqrt2) \right]
\eeq
has ${\cal N}=2$ supersymmetry for any function ${\cal H}$ since
$$
\delta^* {\cal H}(\hat V_1 + i \hat V_2/\sqrt2)  
= {\cal H}^\prime \, \delta^*(\hat V_1 + i \hat V_2 / \sqrt2)
= - \Bigl[ \eta D + \ov\eta\ov D \Bigr] {\cal H}.
$$
As found by Lindstr\"om and Ro\v cek~\cite{LR}, these non-trivial couplings
are generated by solutions of the Laplace equation for variables $\hat V_1$
and $\hat V_2/\sqrt2$. 

Expression (\ref{coupling1}) is sufficient to propagate the physical fields 
of the tensor multiplet. If $h\ne0$ however, it is inconsistent in itself since the highest 
components of both $V_1$ and $V_2$ imply ${\cal H}^\prime=0$ 
by their field equations. Consistency and propagation terms for the fields
in the vector multiplet require to add the vector multiplet
Lagrangian (\ref{LMax}) to eq.~(\ref{coupling1}).

This method for coupling a vector and a tensor supermultiplet corresponds to a ${\cal N}=2$
``St\"uckelberg gauging". The simplest example is ${\cal H}(x)= - \frac{1}{2}x^2$, for 
which the Lagrangian (\ref{coupling1}) is
\beq
\label{}
\Dint\Bigl[  \frac{1}{2}(\phi+\ov\phi - hV_2)^2 - (L - hV_1)^2 \Bigr].
\eeq
It includes the free tensor multiplet Lagrangian
$\Dint \Bigl[ \phi\ov\phi - L^2 \Bigr]$ and mass terms for $V_1$
and $V_2$: there is a gauge where $L$ and $\phi$ are eliminated and 
$V_1$ and $V_2$ acquire a mass proportional to $h$. 

Two particular choices for the function ${\cal H}$ lead to terms of special
interest. Firstly, the Fayet-Iliopoulos terms (\ref{FI2}) are obtained from
a linear function ${\cal H}(x) = \xi x / h$, with $\xi$ complex. Secondly, 
choosing ${\cal H}(x) =  \frac{i}{2\sqrt2 h} \,\zeta\, x^2$ ($\zeta$
real) leads to the action contribution
\beq
\label{BFterm}
\begin{array}{rcl}
{\cal L}_{BF} &=&
\zeta\, \bigint d^4 x\Dint \left[ LV_2 + (\phi+\ov\phi)V_1 - h V_1V_2 \right]
\crbig
&=& -\frac{1}{2}\zeta\Fint [\phi X + 2\,\chi^\alpha W_\alpha]
-\frac{1}{2}\zeta\Fbarint [\ov\phi \ov X + 2\,\ov\chi_\dalpha \ov W^\dalpha]
-\zeta h \Dint V_1V_2 .
\end{array}
\eeq
It is important to notice that this expression has a smooth $h\rightarrow0$ 
limit: the resulting contribution
$$
\zeta\, \bigint d^4 x\Dint \left[ LV_2 + (\phi+\ov\phi)V_1\right],
$$
which is the ${\cal N}=2$ extension of the Chern-Simons coupling 
$\epsilon^{\mu\nu\rho\sigma} B_{\mu\nu} F_{\rho\sigma}$,
also exists if the tensor multiplet is not charged under the abelian 
gauge symmetry. 

Hence a gauge-invariant coupling of a tensor multiplet $(L, \phi)$ 
to the vector multiplet $(V_1,V_2)$ can be described by
the Lagrangian
\beq
\label{newN=2int}
{\cal L}_{vect.-tens.} = \Dint \left[ {\cal H}(\hat V_1 + i\hat V_2/\sqrt2 ) 
+{\cal H}^*(\hat V_1 - i\hat V_2/\sqrt2) \right] + {\cal L}_{\cal F} \, ,
\eeq
where ${\cal L}_{\cal F}$ is the vector multiplet action with prepotential
${\cal F}$. The Chern-Simons and Fayet-Iliopoulos terms can be added
if the tensor multiplet does not transform under the gauge symmetry. On the other hand, they
arise from ${\cal H}$ if the tensor multiplet is charged. 

% Section 3

\section{Nonlinear \boldmath{${\cal N}=2$} supersymmetry and the Born-Infeld theory} 
\label{secN=2nonlin}

\subsection{Partially broken supersymmetry and a nonlinear deformation} 
\label{secpartial}

In our formulation of ${\cal N}=2$ supersymmetry on ${\cal N}=1$ superspace, partial 
supersymmetry breaking corresponds to a simple nonlinear deformation of 
the linear supersymmetry transformations.
Suppose then that instead of transformations (\ref{*vector1}) we use 
\beq
\label{*vectornlin}
\begin{array}{l}
\delta^*X = \sqrt 2 \, i \, \eta^\alpha W_\alpha ,
\qquad\qquad\qquad\qquad
\delta^*\ov X = \sqrt 2 \, i \, \ov\eta_\dalpha \ov W^\dalpha ,
\crbig
\delta^* W_\alpha =  \sqrt 2 \, i \left[ \frac{1}{2\kappa} u \, \eta_\alpha 
+ \frac{1}{4}\eta_\alpha \ov{DD}\,\ov X 
+ i (\sigma^\mu\ov\eta)_\alpha \, \partial_\mu X \right] ,
\crbig
\delta^* \ov W_\dalpha = \sqrt 2 \, i \, \left[ \frac{1}{2\kappa} \ov u \, \ov\eta_\dalpha
+ \frac{1}{4} \ov\eta_\dalpha {DD}\, X 
- i (\eta\sigma^\mu)_\dalpha \, \partial_\mu \ov X \right] ,
\end{array}
\eeq
with an arbitrary (nonzero) constant $\kappa$ with dimension $(length)^2$ and a 
complex phase $u$, $|u|=1$. This modification does not affect the second 
supersymmetry algebra or the Bianchi identity verified by $W_\alpha$. In this
nonlinear variation, the gaugino in $W_\alpha$ transforms like a Goldstino. 

With modified transformations (\ref{*vectornlin}), the second supersymmetry 
variation of the Lagrangian (\ref{LMax}) acquires the new contribution
$$
\frac{\sqrt2i}{4\kappa} u \Fint {\cal F}^{\prime\prime}(X/m) W\eta 
+ {\rm h.c.}
= \frac{m}{4\kappa} u\, \delta^*\Fint {\cal F}^\prime (X/m) + {\rm h.c.}
$$
Hence, the modified Lagrangian 
\beq
\label{LMax3}
\begin{array}{rcl}
{\cal L}_{\cal F} &=& 
\frac{1}{4}\Fint \Bigl[ {\cal F}^{\prime\prime}(X/m)WW 
-\frac{m}{2}{\cal F}^\prime(X/m) \ov{DD}\ov X  - \frac{m}{\kappa}
u\, {\cal F}^\prime (X/m)
- \frac{1}{2} \xi_1 X \Bigr] + {\rm h.c.}
\crbig
&& + \xi_2 \Dint V_2
\end{array}
\eeq
has linear ${\cal N}=1$ supersymmetry and a second, nonlinearly-realized,
supersymmetry with variations (\ref{*vectornlin}). The introduction of the terms 
with coefficient $\kappa^{-1}$ breaks then spontaneously ${\cal N}=2$. 
The resulting superpotential is
\beq
\label{wis}
w = -\frac{m}{4\kappa} u \, {\cal F}^\prime(X/m) - \frac{1}{8}\xi_1X.
\eeq
It includes a new ``magnetic" term proportional to the first derivative of the prepotential~\cite{APT, Fujiwara:2004kc}. 
Together with the Fayet-Iliopoulos term for $V_2$, this superpotential 
leads to the scalar potential
\beq
\label{Vis}
V = (\Re{\cal F}^{\prime\prime})^{-1}\left( \frac{1}{8}\xi_2^2 
+ \frac{1}{16} \left| \frac{1}{2} \xi_1 + \frac{1}{\kappa} u\, {\cal F}^{\prime\prime} \right|^2
\right).
\eeq
It has a stationary point with
$$
\begin{array}{l}
\Re {\cal F}^{\prime\prime} = \left[ 2\xi_2^2\kappa^2 + \frac{1}{4}\xi_1^2\kappa^2
(\Re u)^2 \right]^{1/2},
\qquad\qquad
\Im {\cal F}^{\prime\prime} = \frac{1}{2}\xi_1\kappa \, \Im u,
\crbig
V = \frac{1}{8\kappa^2} \Re{\cal F}^{\prime\prime}
+ \frac{\xi_1}{16\kappa} \Re u.
\end{array}
$$

The nonlinear second supersymmetry transformations (\ref{*vectornlin}) of
$W_\alpha$ translates into a modified variation of the vector superfield $V_2$:
\beq
\label{*V2nlin}
\begin{array}{rcl}
\delta^* V_2  &=& \sqrt2 i \Bigl[ \eta D + \ov\eta\ov D \Bigr] V_1
+ \frac{i}{\sqrt2\kappa} \, u \,\ov{\theta\theta}\theta\eta
- \frac{i}{\sqrt2\kappa} \, \ov u \, \theta\theta \ov{\theta\eta}
\crbig
&=& \sqrt2 i \Bigl[ \eta D + \ov\eta\ov D \Bigr] \Bigl(V_1 + \frac{1}{4\kappa}
\Re u \, \theta\theta\ov\theta\ov\theta  \Bigr)
- \frac{1}{\sqrt2\kappa} \Im u \Bigl( \ov{\theta\theta}\theta\eta +
\theta\theta \ov{\theta\eta} \Bigr).
\end{array}
\eeq
Notice that the nonlinear deformation does not affect the existence of
Fayet-Iliopoulos terms: only fermion variations are modified.  

Note that the phase $u$ is in principle an arbitrary parameter.
By imposing the exchange symmetry $\lambda\leftrightarrow\psi$ 
of the two gaugino mass terms\footnote{The four-fermion interactions 
do not depend on the superpotential.}, $u$ is found to be purely imaginary, 
$u=\pm i$~\cite{APT}.

\subsection{The nonlinear constraint}

The Born-Infeld theory with linear ${\cal N}=1$ supersymmetry can be nicely derived
as a nonlinear realization of ${\cal N}=2$ supersymmetry on ${\cal N}=1$ superspace. This
nonlinear realization can be obtained by imposing a nonlinear constraint
on the linear ${\cal N}=2$ vector multiplet introduced in the previous section. 

Following Ro\v cek and Tseytlin~\cite{RT, Rocek}, we construct the nonlinear
realization by imposing the constraint
\beq
\label{*constraint}
\frac{1 }{ \kappa} X = WW - \frac{1}{2} X \ov{DD}\ov X,
\eeq
where $\kappa$ describes the scale of supersymmetry breaking and has dimension ({\emph{length}})$^2$. Since 
$$
X = \frac{WW}{ \kappa^{-1} + \frac{1}{2} \ov{DD}\ov X }
$$
and $W_\alpha W_\beta W_\gamma =0$, the constraint implies $W_\alpha X =0$,
and the derivative in the variation (\ref{*susy1}) disappears. The right-hand side of
constraint (\ref{*constraint}) is then invariant under the linear second supersymmetry 
(\ref{*vector1}) and covariance of the constraint requires a nonlinear modification
of $\delta^*W_\alpha$, given by eq.~(\ref{*vectornlin}) with the phase $u=1$:
\beq
\label{*vector2}
\begin{array}{l}
\delta^*X = \sqrt 2 \, i \, \eta^\alpha W_\alpha ,
\qquad\qquad\qquad\qquad
\delta^*\ov X = \sqrt 2 \, i \, \ov\eta_\dalpha \ov W^\dalpha ,
\crbig
\delta^* W_\alpha = \sqrt 2 \, i \left[ \frac{1}{2\kappa}\eta_\alpha 
+ \frac{1}{4}\eta_\alpha \ov{DD}\,\ov X 
+ i (\sigma^\mu\ov\eta)_\alpha \, \partial_\mu X \right] ,
\crbig
\delta^* \ov W_\dalpha = \sqrt 2 \, i \, \left[  \frac{1}{2\kappa}\ov\eta_\dalpha
+ \frac{1}{4} \ov\eta_\dalpha {DD}\, X 
- i (\eta\sigma^\mu)_\dalpha \, \partial_\mu \ov X \right] .
\end{array}
\eeq
The superfield $W_\alpha$ includes then the Goldstino of the nonlinear 
supersymmetry with transformations $\delta^*$.
With this modification,
\beq
\label{*modif}
\delta^* \left[ WW - \frac{1}{2} X \ov{DD}\ov X \right] = \frac{\sqrt 2 i}{ \kappa}\, \eta W = 
\frac{1}{ \kappa}\,\delta^* X,
\eeq
as required by the constraint (\ref{*constraint}). 
Since $\delta^*\eta_\alpha$ and $D_\alpha\eta_\beta$ vanish, the modification does not affect 
the algebra and the Bianchi identity. Eqs.~(\ref{*vector2}) and (\ref{*constraint}) 
provide a nonlinear realization of ${\cal N}=2$ supersymmetry, with linearly-realized
${\cal N}=1$. 

The constraint (\ref{*constraint}) allows to express $X$ as a function of $WW$ and 
of its supersymmetric derivatives (see below), and the ${\cal N}=2$ super-Maxwell theory
reduces simply to the Fayet-Iliopoulos term
\beq
\label{*Maxwell2}
{\cal L}_{Max.} = \frac{1}{4\kappa g^2 } \Fint X  +\frac{1}{ 4\kappa g^2} \Fbarint \ov X,
\eeq
adding a constant gauge coupling $g$ (which is useless in the abelian theory 
without charged fields). 
The invariance of this action follows from
\begin{eqnarray}
\int d^4x \, \delta^* {\cal L}_{Max.} &=&  
- \frac{ \sqrt 2 i }{16 \kappa g^2} \int d^4x \Fint \, DD (\eta W) + {\rm h.c.}\nonumber
\crbig
&=& \frac{\sqrt2 }{ 4\kappa g^2}  \int d^4x \Fint\, \partial_\mu(\eta\sigma^\mu W) + {\rm h.c.}\nonumber
\end{eqnarray}
Solving the constraint and expanding the action in components shows that
theory (\ref{*Maxwell2}) provides an extension of the Born-Infeld Lagrangian
with linear ${\cal N}=1$ supersymmetry and a nonlinearly-realized second 
supersymmetry with variations (\ref{*vector2})~\cite{BG}. 

\subsection{The \boldmath{${\cal N}=2$} Born-Infeld theory} \label{subsec:BI}

Bagger and Galperin~\cite{BG} have shown how to solve the constraint 
(\ref{*constraint}). The result is
\beq
\label{Xis}
X = \kappa W^2 -\kappa^3 \ov{DD} \left[ \frac{ W^2\ov{W^2} }{ 
1+ A + \sqrt{1 + 2A + B^2 }} \right] ,
\eeq
where 
$$
A = \frac{\kappa^2}{2 } ( DD\, W^2  + \ov{DD} \,\ov W^2) = A^*, \qquad\qquad
B = \frac{\kappa^2}{2 } ( DD\, W^2  - \ov{DD} \,\ov W^2) = -B^*.
$$
The derivation is also given in Appendix~\ref{A3}.
Notice that
\beq
\label{ABis}
\begin{array}{rcl}
\left. A \right|_{\theta=0} &=& -8\kappa^{2} \left[ -\frac{1}{4}F_{\mu\nu}F^{\mu\nu} + \frac{1}{2}d^2
+ \frac{i}{2}\lambda\sigma^\mu\partial_\mu\ov\lambda
-\frac{i}{2}\partial_\mu\lambda\sigma^\mu\ov\lambda \right] ,
\crbig
\left. B \right|_{\theta=0} &=& 4 i \kappa^{2} 
\left[ \frac{1}{2}F_{\mu\nu}\tilde F^{\mu\nu} - \partial_\mu
( \lambda\sigma^\mu\ov\lambda ) \right],
\qquad\qquad
F_{\mu\nu} \tilde F^{\mu\nu} = \partial_{\mu}(\epsilon^{\mu\nu\rho\sigma}A_\nu 
F_{\rho\sigma}) .
\end{array}
\eeq
The expression (\ref{Xis}) has been obtained by Cecotti and Ferrara~\cite{CF} in the
context of ${\cal N}=1$ superspace, with however an ambiguity in the ${\cal N}=1$ 
supersymmetrization which is removed when the second (nonlinear) supersymmetry 
is imposed~\cite{BG}.
Notice also that, as expected, the solution (\ref{Xis}) is compatible with $X=\frac{1}{2}\ov{DD}V_1$, but it only defines $V_1$ up to a linear superfield. 

The Lagrangian (\ref{*Maxwell2}) includes the following gauge kinetic terms (see Appendix~\ref{A3}):
\beq
\label{BI}
\begin{array}{rl}
{\cal L}_{Max.} \quad\longrightarrow\quad&
\frac{1}{8\kappa^2g^2}  \left[ 1 - \sqrt{ 1 + 4\kappa^{2}F_{\mu\nu}F^{\mu\nu} 
- 4\kappa^{4}(F_{\mu\nu}\tilde F^{\mu\nu})^2}  \right]
\crbig
&= -\frac{1}{4g^2} F_{\mu\nu}F^{\mu\nu} + {\cal O}(\kappa^{2}{F_{\mu\nu}}^4)
\crbig
&= \frac{1}{8\kappa^2g^2} \left[ 1 - \sqrt{ - {\rm det}(\eta_{\mu\nu} + 2\sqrt2 \,\kappa
F_{\mu\nu})} \right].
\end{array}
\eeq
It also includes terms with derivatives of $F_{\mu\nu}$, fermionic contributions
and the following auxiliary contribution:
\beq
\label{BIdterms}
{\cal L}_{Max.} \quad\longrightarrow\quad
\frac{1}{8\kappa^2g^2} \left[ 1 - \sqrt{1 - 8 \kappa^{2}d^2} \right] 
= \frac{1}{2g^2}d^2 + {\cal O}(d^4) .
\eeq

The usual Born-Infeld Lagrangian for a $D3$ brane is
\beq
\label{BI2}
{\cal L}_{BI} = -T_3\frac{2\pi}{ g^2} \sqrt{-{\rm det}(\eta_{\mu\nu} + 2\pi \alpha'  F_{\mu\nu})} ,
\eeq
where $g$ is the gauge coupling and $T_3$ is the brane tension:
\beq
\label{T3is}
T_3 = \frac{1}{ (2\pi)^{3}({\alpha^\prime})^{2}}\, .
\eeq
Comparing then eqs.~(\ref{BI}) and (\ref{BI2}) leads to the identifications~\cite{Antoniadis:2004uk}:
\beq
\label{kappais}
T_3 = \frac{1}{ 16\pi\kappa^2} \qquad ;\qquad
\kappa = \frac{ \pi\alpha^\prime }{  \sqrt2}.
\eeq

Note that, upon imposing the nonlinear constraint (\ref{*constraint}), the DBI action is identical to 
the first FI--term proportional to $\xi_1$ in eq.~(\ref{FI2}). On the other hand, the second FI--term 
proportional to $\xi_2$ is obviously invariant under the nonlinear supersymmetry transformation 
(\ref{*V2nlin}) (with $u=1$), compatible with the constraint, and can be also added to the action. 
This leads to an additional contribution, linear in the D-term auxiliary $d$.

%%%% Section 4

\section{The \boldmath{${\cal N}=2$} dilaton} \label{secN=2sugra}

In string theory, the gauge coupling is related to the VEV of the dilaton field and the contribution (\ref{BIdterms}) provides a dilaton potential at the level of the disk world-sheet topology. On the other hand, a tree-level dilaton potential at the level of spherical topology can be generated by going off-criticality, away from ten dimensions. This will be used in the next section, where we study supersymmetry breaking vacua with dilaton stabilization. In the context of the effective field theory, a non-critical dilaton potential can be described as a gauging of the ${\cal N}=2$ supergravity of the closed string sector, that we present in this section.

In type IIB superstrings compactified to four dimensions on a Calabi-Yau threefold, the dilaton 
belongs to a hypermultiplet of four-dimensional ${\cal N}=2$ supergravity. This multiplet can 
be dualized into a single-tensor or a double-tensor multiplet since two of its four scalar 
components are actually modes of the NS-NS (Neveu-Schwarz) and R-R (Ramond-Ramond)
two-form fields. Taking the orientifold (with $D9$ branes) of this
theory leads to type I strings compactified on a Calabi-Yau space, the axion partner of the dilaton 
being a mode of the R-R two-form field of the original type IIB theory.

To describe the non-critical dilaton potential which we will use in the next section, we use ${\cal N}=2$ supergravity with a single hypermultiplet
and we gauge the axion shift symmetry using the graviphoton as gauge field. The quaternionic 
scalar manifold is $SU(2,1)/SU(2)\times U(1)$, which is also a K\"ahler manifold.
The terms of the ${\cal N}=2$ theory relevant to our purposes are simply~\cite{N=2sugra}:
\beq
e^{-1}{\cal L}_{N=2} = - \frac{M_P^2}{2} R +  M_P^2 \, h_{ab}(q)\,
 (\partial_\mu q^a)(\partial^\mu q^b) - {\cal V} (q) + \ldots \,,
\eeq
where $q^a$, $a=0,1,2,3$, are the four hypermultiplet real scalar fields and $h_{ab}(q)$
is the $SU(2,1)/SU(2)\times U(1)$ metric. Here, $R$ is the scalar curvature and $M_P$ the reduced Planck mass. Defining precisely the scalar potential 
${\cal V} (q)$ requires some preliminaries. 

The metric $h_{ab}(q)$ of a quaternionic manifold is hermitian with respect to a triplet 
of complex structures $J^x$ verifying the quaternionic algebra\footnote{We follow the conventions
of Ref. \cite{D'AF}. See in particular the Appendix.}
$$
J^x J^y = - \, \delta^{xy}\, I + \epsilon^{xyz}J^z, \qquad\qquad (x,y,z=1,2,3).
$$
The three {\it hyper-K\"ahler forms} $K^x_{ab} = h_{ac}(J^x)^c_b$
are then covariantly closed with respect to a $SU(2)$ connection $\omega^x$:
\beq
\label{Kxclosed}
d \, K^x + \epsilon^{xyz} \, \omega^y \wedge K^z =0 \, .
\eeq
In the case of a quaternionic manifold, the $SU(2)$ curvatures are proportional to 
the hyper-K\"ahler forms,
\beq
\label{SU(2)curv}
\Omega^x = \frac{1}{2}\,\Omega^x_{ab} \, dq^a \wedge dq^b
\equiv d \, \omega^x + \frac{1}{2} \epsilon^{xyz} \, \omega^y \wedge \omega^z
= \lambda \, K^x, \qquad\qquad (\lambda\ne0).
\eeq
This is the case relevant to hypermultiplets coupled to ${\cal N}=2$ 
supergravity.\footnote{In the case of global ${\cal N}=2$ supersymmetry, 
the $SU(2)$ curvature $\Omega^x$ vanishes and the manifold is hyper-K\"ahler.}
Eqs.~(\ref{Kxclosed}) and (\ref{SU(2)curv}) can be viewed as a set of equations for 
the connection $\omega^x_{ab}$. The definitions of $J^x$ and $K^x$ and the
proportionality equation (\ref{SU(2)curv}) lead to
\beq
\label{curvmetric}
h^{cd} \, \Omega_{ac}^x \, \Omega_{db}^y = -\lambda^2 \, \delta^{xy} \, h_{ab}
+ \lambda \, \epsilon^{xyz} \, \Omega_{ab}^z \, ,
\eeq
where $h^{cd}(q)$ is the inverse quaternionic metric. This equation defines 
$\lambda h_{ab}$ in terms of the curvatures $\Omega^x$, or in terms of the connections 
$\omega^x$. Supersymmetry leads in general to kinetic terms (in $M_P$ mass units):
$$
-\frac{1}{2} \, eR - e \,\lambda\, h_{ab}(q) (\partial_\mu q^a)(\partial^\mu q^b) 
+ \ldots
$$
and canonical normalization is obtained with the choice $\lambda=-1$. 

With one hypermultiplet only, the 
theory has a single vector field, the graviphoton, and we can gauge one isometry of the quaternionic metric.
The Killing vector $k^a$ for the gauged symmetry is defined from the symmetry action
\beq
q^a \,\rightarrow\, q^a + \epsilon \, k^a (q)\, ,
\label{killvec}
\eeq
where $\epsilon$ is the (infinitesimal) parameter of the transformation. 
Each Killing vector of a quaternionic manifold can be expressed in terms of a triplet 
of prepotentials ${\cal P}^x$:
\beq
\label{Pxcond}
2 \, k^a \, \Omega_{ab}^x = - ( \partial_b{\cal P}^x + \epsilon^{xyz} \, \omega^y_b \, {\cal P}^z )
\equiv  - \nabla_b{\cal P}^x \, .
\eeq
Eq.~(\ref{curvmetric}) leads then to:
\beq
\label{Pxmetric}
k^a = \frac{1}{6\lambda^2} \, \sum_{x=1}^3 h^{bc} \, \partial_b ( {\cal P}^x \,
\Omega^x_{cd}) \, h^{da}.
\eeq

For our $SU(2,1)/SU(2)\times U(1)$ manifold, an appropriate triplet of $SU(2)$
connections with closed curvatures for any value of the real parameter $r$ is~\cite{N=2D=5}
\beq
\label{SU(2)connec}
\omega^1 = - r \frac{d\tau }{ \sqrt V} \, ,
\qquad
\omega^2 = r \frac{d\theta }{ \sqrt V} \, ,
\qquad
\omega^3 = -\frac{r}{4V} ( d\sigma - 2\tau d\theta + 2\theta d\tau),
\eeq
using the basis $q^a = (V, \sigma, \theta, \tau)$. In this basis, gauging the 
axionic shift symmetry on $\sigma$ means that we choose a Killing vector (\ref{killvec})
\beq
\label{Killing}
k^a = ( 0, 1, 0, 0 ),
\eeq
{\it i.e.} we gauge the transformation $\sigma \, \rightarrow \, \sigma + \epsilon$. 
With these connections, 
we choose the prepotential triplet
\beq
\label{Pxis}
\vec {\cal P} = \Bigl( \, 0 \, , \, 0 \, , \, -\frac{1}{ V} \, \Bigr)
\eeq
and equation (\ref{Pxcond}) is verified if $r=2$ in expressions (\ref{SU(2)connec}).
We can then use eq.~(\ref{Pxmetric}) or eq.~(\ref{curvmetric}) to obtain the quaternionic
metric, choosing $\lambda=-1$ as required by supersymmetry and canonical normalization 
of kinetic terms. The result is then
\beq
\label{quatmetric}
ds^2 = \frac{dV^2}{ 2V^2} +\frac{1}{ 2V^2}(d\sigma - 2\tau d\theta + 2\theta d\tau )^2
+\frac{2}{ V}(d\theta^2 + d\tau^2)
\equiv h_{ab}(q) \, dq^a dq^b \, .
\eeq
Since the scalar manifold is also K\"ahler, it can be derived from the K\"ahler potential 
\beq
\label{SU2Kis}
K = - 2 \ln(S + \ov S - 2 C\ov C) \,,
\eeq
with definitions 
\beq
\label{SandCare}
\begin{array}{rcl}
S &=& V + \theta^2 + \tau^2 + i \sigma, 
\crbig
C &=& \theta - i \tau\, .
\end{array}
\eeq
Since $S + \ov S - 2 C\ov C = 2V$, $V$ is the four-dimensional  dilaton field associated to the four-dimensional type II string coupling $e^{\varphi_4}$: $V=e^{-\varphi_4}$ (see next section).
Moreover, the shift isometry on $\sigma$ (\ref{killvec}) follows by its (Poincar\'e) dualization from the NS-NS antisymmetric tensor.

Gauging symmetries generates in particular a scalar potential. 
For a single hypermultiplet, this potential receives two contributions~\cite{N=2sugra}:
\beq
\label{potV=2}
{\cal V}(q) = {g_*^2 M_P^4} \left[ 4 \, h_{ab}(q) \, k^a k^b 
- 3 \sum_{x=1}^3 {\cal P}^x{\cal P}^x  \right] L^0 \ov L^0 \, ,
\eeq
where $g_*$ is the gauge coupling. The ``section" $L^0$ can be chosen $L^0=1$ since vector 
multiplets are absent. With our Killing vector $k^a$, our prepotentials ${\cal P}^x$ and our metric 
(in particular with $h_{11}= (2V^2)^{-1}$) the scalar potential is
\beq
\label{dilatonpot}
{\cal V}(q) = {g_*^2 M_P^4} \left[ 4 \, h_{11}(q) 
- \frac{3}{ V^2}  \right] = - \frac{g_*^2 M_P^4}{ V^2} \, .
\eeq

%%%% Section 5

\section{Scalar potential} \label{secpotential}

We are now interested in the disk  contributions to the four-dimensional scalar potential induced by the presence of magnetized branes in a type I orientifold compactification. It will be shown that it receives three independent contributions. First, the uncancelled NS-NS tadpole contribution is encoded in the  ``branes tension deficit" $\delta T$ which gives the tree-level dilaton tadpole. The second contribution comes from the ``anomalous" FI-term proportional to $\xi$, while the last term arises from the supersymmetrization of the DBI action, as presented in the previous section. The only consistent vacuum formed by magnetic fluxes turns out to be supersymmetric, where at least one of the closed string moduli (the dilaton) remains unfixed. The situation may however be different when strings propagate in a non-trivial background: either in the presence of three-form closed string fluxes, or in non-critical dimensionality. In this case, an extra contribution to the scalar potential from the sphere world-sheet changes the equation of motion for the dilaton and may lead to different vacua with broken supersymmetry in curved space-time. 

\subsection{FI-terms from magnetic fluxes}

Let us consider type I string theory compactified on a Calabi-Yau threefold ${\cal M}_6$. The ${\cal N}=1$, $d=4$ action contains a number of scalar fields describing the size $J_\alpha$ and shape $\tau_k$ moduli of the internal manifold as well as the dilaton field $\varphi$. In addition to them, there exist $h_{1,1}+1$ axionic fields $C_{\alpha}$ and $C_0$ which arise from the compactification of the 
ten-dimensional two-form $C_2$. Their kinetic terms may be diagonalized in terms of the chiral superfields $T_\alpha$,  $S_I$ and $U_k$ as~\cite{Antoniadis:1996vw, Grimm:2004uq}
\beq
S_I = e^{-\varphi}\frac{V_6}{ (4\pi^2\alpha')^3} +i\, c \, , \qquad T_\alpha = -e^{-\varphi}
\frac{J_\alpha}{ 4\pi^2\alpha'} +i \, c_\alpha \, ,
\label{STU}
\eeq
where the overall volume $V_6$ is defined by the integral of the K\"ahler moduli $J = J_\alpha \omega^\alpha$ over the internal manifold ${\cal M}_6$: $V_6 =\int_{{\cal M}_6} J\wedge J \wedge J$, and $\{\omega_\alpha\}$ is a basis of the two-forms on ${\cal M}_6$. Note that the type I dilaton superfield $S_I$ differs from the one that appears in the universal type II hypermultiplet (\ref{SandCare}).
Their K\"ahler potential in the absence of fluxes is
\beq
{\cal K} = -M_P^2\left[ \ln \left(S_I+\ov{S}_I\right) + \ln \int_{{\cal M}_6}\left(T+\ov{T}\right) \wedge 
\left(T+\ov{T}\right) \wedge\left(T+\ov{T}\right) \right]\, , \quad \quad T = T_\alpha \omega^\alpha .
\label{sc:K1}
\eeq
For simplicity, we omit the complex structure moduli $U_k$ from our discussion.

Let us now consider  $K$ $U(1)_a$ magnetized $D9$ branes, with $a=1,\dots, K$. These give rise 
to $K$ gauge fields with non-trivial gauge bundle on ${\cal M}_6$. Let us denote the corresponding 
gauge superfields by $V_a$. The gauge bundles on the internal manifold manifest themselves  by 
topological couplings $Q_0^a$ and $Q_\alpha^a$ of the axionic fields $c_0$ and $c_\alpha$ to the 
corresponding $U(1)_a$ field strengths.  These can be determined by the dimensional reduction of 
the Wess-Zumino action:
\beq
Q^a_0 =\frac{1}{ (2\pi)^3}\int_{{\cal M}_6} F^a\wedge F^a \wedge F^a \, ,
\qquad\qquad Q^a_\alpha = \frac{1}{ 2\pi} F^a_\alpha  \, ,
\eeq 
where the $F^a_\alpha$'s are the (quantized) components of the $U(1)_a$ field strengths along the 
two-cycle $\alpha$. In other words, these fluxes  modify  the K\"ahler potential (\ref{sc:K1}) to
\beq
\frac{1}{ M_{P}^2}{\cal K} = -\log \left(S_I+\ov{S}_I + \sum_a Q^a_0V_a\right) - \log{ \int_{{\cal M}_6}}
\left( T + \ov T +\sum_aQ^a V_a\right)^3 \, ,
\label{sc:K2}
\eeq
where $Q^a = Q^a_\alpha \omega^\alpha$ and the power $3$ is defined in terms of the wedge product as in (\ref{sc:K1}).

In addition to the topological coupling, one is able to extract from  the K\"ahler potential ${\cal K}$ 
the FI-term $\xi_a$ induced by the magnetic fluxes for each $U(1)$ gauge 
component~\cite{Dine:1987xk, Atick:1987gy,  Cremades:2002te}:
\beq
\Dint \,  {\cal K} = \left(\frac{\partial {\cal K} }{ \partial V_a}\right)_{V_a=0, \, \theta=0} \Dint \, V_a + \cdots
\eeq
or, equivalently,
\beq
\label{xiST}
\begin{array}{rcl} \displaystyle
\frac{\xi_a }{ g_a^2}  &\equiv& - \displaystyle \left(\frac{\partial {\cal K} }{ \partial V_a}\right)_{V_a=0, \,\theta=0}  
= M_{P}^2 \left( \frac{Q^a_0 }{ S_I+\bar{S}_I} + \frac{1 }{ 2\int_{{\cal M }_6}({\rm Re}T)^3}\int_{{\cal M}_6} 
Q^a \wedge {\rm Re}T\wedge {\rm Re}T \right) \, 
\crbig
&=& \displaystyle e^{-\varphi} \frac{ M_s^2 }{ 2\pi}\left(\frac{1}{ (4\pi^2\alpha')^3} \int\left( {\cal F}\wedge {\cal F}\wedge {\cal F} -  \, J \wedge J \wedge {\cal F}  \right) \right) \,   
\crbig
&\equiv & \displaystyle e^{-\varphi} \frac{ M_s^2 }{ 2\pi} \frac{\bar{\xi}_a }{ g_a^2} \,\, , \qquad\qquad
\frac{\bar{\xi}_a }{ g_a^2}=\frac{1}{ (4\pi^2\alpha')^3} \int\left( {\cal F}\wedge {\cal F}\wedge {\cal F} - 
 \, J \wedge J \wedge {\cal F}  \right),
\end{array}
\eeq
where $M_s=(\alpha')^{-1/2}$, ${\cal F}^a=2\pi\alpha'F^a$ and we have used the definitions 
(\ref{STU}) and the relation for the reduced Planck mass: 
\beq
\label{MP}
M_{P}^2 = \frac{1}{ \pi}M_s^2\, e^{-2 \varphi}   \frac{V_6}{ (4\pi^2\alpha')^3}\, .
\eeq

\subsection{Scalar potential}

On the world-volume of each $D9$ brane stack lives a $U(N_a)$ gauge theory. Let us restrict to its $U(1)_a$ subgroup whose NS-NS sector   is described at low energy by the ten-dimensional DBI action
\beq
\label{sec:DB;BI1}
S_{BI,a} = -T_9 \int_{\Sigma_a} d^{10}x \, e^{-\varphi} \, \sqrt{ -{\rm det} (G+2\pi \alpha' F_a)} \; , \qquad T_p = \frac{1 }{ (2\pi)^p \, {\alpha'\,}^\frac{p+1}{2}} \, ,
\eeq
where $\Sigma_a$ is the ten-dimensional world-volume of the a-th $D9$ brane with metric $G$ and $F_a$ is the field strength of the $U(1)$ gauge theory. 
In the reduction relevant for us, the action (\ref{sec:DB;BI1}) simplifies to 
\beq
\label{sec:DB;BI2}
S_{BI,a} = -T_9 \int_{{\cal M}_4} d^4x \, e^{-\varphi} \, \sqrt{ -{\rm det} (G_{4}+2\pi \alpha' F_{4,a})} \,  \cdot \int_{{\cal M}_6}d^6y \, \sqrt{ {\rm det} (G_{6}+2\pi \alpha' F_{6,a})} \, ,
\eeq
where $G_4$, $G_6$ and $F_{4,a}$, $F_{6,a}$ are the metric and $U(1)$ field strength on ${\cal M}_4$ and  ${\cal M}_6$ respectively. Let us consider the case where the metric moduli of $G_6$ are stabilized at specific points on their moduli space. 
The last factor of the action (\ref{sec:DB;BI2}) can then be considered as  constant from a 
four-dimensional viewpoint:
\begin{eqnarray}
\label{sec:DB;BI3}
S_{BI,a} &=& -T_3 \, \frac{2\pi }{ g^{2}_a} \int_{{\cal M}_4} d^4x \, e^{-\varphi} \, \sqrt{ -{\rm det} (G_{4}+2\pi \alpha' F_{4,a})}  \, ,\\
\frac{2\pi }{ g^{2}_a}&=& \frac{1 }{ (4\pi^2 \alpha')^3}\int_{{\cal M}^6}d^6y \, \sqrt{ {\rm det} (G_{6}+2\pi \alpha' F_{6,a})}\, .\nonumber
\end{eqnarray}
As shown in Section~\ref{secN=2nonlin}, the supersymmetrization of the action (\ref{sec:DB;BI3}) leads to a potential for the auxiliary component $d_a$ of the a-th $U(1)$ gauge superfield. Using eqs.~(\ref{BIdterms}) and (\ref{sec:DB;BI3}), this reads:
\beq
\label{sec:DB;SP}
S_{SP,a} = -T_3 \, \frac{ 2\pi }{ g^2_a} \int_{{\cal M}_4} d^4x \, \sqrt{-{\rm det}\,G_4} \,
e^{-\varphi} \, \left(1-1+ \sqrt{1-  (2\pi\alpha' d_a)^2} \right)\, .
\eeq
The first term in the bracket comes from the dilaton tadpole contribution, whereas the factor $1- \sqrt{1- (2\pi\alpha' d_a)^2} $ comes from the  DBI action (\ref{BIdterms}). 

Let us now introduce an $O9$ orientifold plane\footnote{Note that the expressions presented here are also valid in the case of $D3/D7$ magnetized branes~\cite{AKM2}.}. It is defined as the set of fixed points of the orientifold projection ${\cal O} = \Omega$, where $\Omega$ is the world-sheet parity. The $O9$ plane is a ten-dimensional object whose effective action is 
\beq
\label{sec:DB;O9}
S_{O9} = -32 \, T_9 \int_{{\cal M}_{10}} d^{10}x \, e^{-\varphi} \, \sqrt{ -{\rm det} \, G} \, .
\eeq
Note that the integral is over the ten-dimensional space-time ${\cal M}_{10} = {\cal M}_{4} \times {\cal M}_{6}$. After compactification to four dimensions, the action (\ref{sec:DB;O9}) reads
\beq
\label{sec:DB;O92}
S_{O9} = -32 \,T_3\, \frac{V_6}{(4\pi^2\alpha')^3} 
\int_{{\cal M}_{4}} d^4x \, e^{-\varphi} \, \sqrt{ -{\rm det} \, G_4} \, , 
\eeq
where  the volume $V_6$ is taken in a particular point on its moduli space determined by the stabilization procedure. 

The various contributions of the K stacks of branes and of the orientifold planes to  the scalar potential arise from  their tensions, the supersymmetrization of the DBI action and  the FI-terms. Using eqs.~(\ref{xiST}), (\ref{sec:DB;SP}) and (\ref{sec:DB;O92}), the potential  then reads (in the string frame)
\begin{eqnarray}
{\cal V}(\varphi, J_\alpha, d_a) &\!\!=\!\!& T_3 \, e^{-\varphi}\left[ \left(\sum_aN_a 
\frac{2\pi}{g^2_a} - 32 \frac{V_6}{(4\pi^2\alpha')^3} \right) \right.
\nonumber \\
&&\left. -   \sum_a N_a \, \frac{2\pi}{ g^{2}_a} \left(1- \sqrt{1- (2\pi\alpha' d_a)^2}\right)  + \sum_a \frac{2\pi}{g^2_a}  (2\pi\alpha' d_a) \, \bar{\xi}_a \right].
\label{sec:DB;V}
\end{eqnarray}
The supersymmetric vacua correspond to points of the moduli space where the VEV's of the auxiliary fields $d_a$ are zero. The expression (\ref{sec:DB;V}) of the potential indicates that $d_a=0$ is only possible if all FI-terms $\xi_a$ vanish, $\xi_a = 0 \, , \, \forall a=1, \dots, K$. Using eq.~(\ref{xiST}), one then obtains the condition: 
\beq
\int\left( {\cal F}_a\wedge {\cal F}_a\wedge {\cal F}_a -  \, J \wedge J \wedge {\cal F}_a   \right) = 0 \, , \qquad \forall a=1,\dots, K\, .
\label{susy1}
\eeq
When these equations are satisfied, the gauge coupling constants $g_a$ defined in eq.~(\ref{sec:DB;BI3}) reduces to the polynomial form  
\beq
g^{-2}_a \sim \frac{1 }{ (4\pi^2\alpha')^3} \int_{{\cal M}_6}\left( J\wedge J\wedge J - J \wedge {\cal F}^a \wedge {\cal F}^a   \right) > 0 \, .
\label{susy2}
\eeq
Note that the dilaton factor $e^{-\varphi}$ was omitted from the definition of the gauge couplings for simplicity. The physical gauge couplings are given by $g_a e^{\langle\varphi\rangle/2}$.

By unitarity, these couplings must  be positive. The conditions (\ref{susy1}) and (\ref{susy2}) are equivalent to the geometrical conditions found in Ref.~\cite{MMMS}. They ensure that the magnetized branes preserve a common  supersymmetry with the orientifold projection. These D-flatness conditions restrict regions of the moduli space where supersymmetry is restored. For given magnetic fluxes ${\cal F}_a$, only particular regions of the K\"ahler moduli space give rise to supersymmetric vacua. 
It should be kept in mind that the above D-flatness conditions are only valid at the point of the open string moduli space where all open string charge states have zero VEV's. Strictly speaking, only a combination of the K\"ahler moduli and charged Higgs-type fields are stabilized by magnetic fluxes.

\subsection{Supersymmetric vacua}

In addition to these necessary conditions (\ref{susy1}) and (\ref{susy2}), a consistent supersymmetric vacuum 
exists only if the sum of the contributions to the R-R  tadpoles vanishes. In the type I compactification, 
these tadpole conditions read
\beq
\sum_a N_a  = 32 \quad {\rm and} \quad \sum_a N_a  \int_{\Pi_\alpha} F_a \wedge F_a = 0 \, , \qquad \forall\alpha = 1, \dots, h_{4}({\cal M}_6)\, ,
\label{tadpole}
\eeq
where $h_4({\cal M}_6)$ is the number of four-cycles $\Pi_\alpha$ (dual to the two-cycle $\alpha$) of the manifold ${\cal M}_6$.
 
When the necessary and sufficient  conditions  (\ref{susy1}), (\ref{susy2}) and (\ref{tadpole}) are satisfied,  
the sum over the brane tensions is exactly compensated by the one of the orientifold planes. The auxiliary fields and the FI-parameters vanish. As it should be, the value of the scalar potential at a supersymmetric vacuum is zero. Note however that the supersymmetric conditions and the R-R tadpole cancellation with only $O9$ planes seem to be incompatible as they stand. One way out to find solutions consists of considering compactifications with orientifold five-planes $O5$. Alternatively, as shown in Ref.~\cite{AKM2}, there exists a second possibility without 
five-planes. Indeed, in the quadratic approximation, the D-flatness condition (\ref{susy1}), which is equivalent to the vanishing of the FI parameter $\xi$ in eq.~(\ref{xiST}), is modified in the presence of charged fields $\phi_i$ to 
\beq
0= \langle d_a \rangle =  \sum_i q^a_i \langle |\phi_i|^2 \rangle + \xi_a\, , 
\eeq
while the tadpole conditions (\ref{tadpole}) remain intact. It is then possible to obtain consistent supersymmetric vacua ($d_a=0\, , \, \forall a$) with non-vanishing FI-parameters $\xi_a \neq 0$. The presence of small non-vanishing VEV's, $\langle |\phi_i|^2 \rangle = v_i^2 \ll M_s^2$, for some charged fields compensates the contribution of the FI-parameter to the D-term.\footnote{The smallness condition for the charged field VEV's guarantees the validity of the perturbative in $\alpha'$ approach.} The sum over the tensions and the scalar potential vanishes. In this way,  it is possible to fix combinations of K\"ahler and  open string moduli in a Minkowski vacuum. 

Since the computation of the FI-term is restricted to the disk amplitude, the dilaton enters only as an overall factor in the scalar potential. It is not constrained by the magnetic fluxes and remains therefore a flat direction. The same conclusion may be drawn from an analysis of the St\"uckelberg couplings. In fact, the stabilized K\"ahler moduli must enter in massive ${\cal N}=1$ vector supermultiplets. This is achieved due to the St\"uckelberg couplings which allow the corresponding R-R axions to become the longitudinal polarizations of the magnetized $U(1)$ gauge bosons. The massive scalar modulus and vector then form the bosonic content of a massive ${\cal N}=1$ vector multiplet. However, in a configuration where all branes satisfy the supersymmetry condition (\ref{susy1}), the maximal rank of the matrix of topological couplings $M_{a\bar{\alpha}} = Q^a_{\bar{\alpha}}$ is given by the number of K\"ahler moduli $h_{1,1}({\cal M}_6)$, (for $\bar{\alpha} = 0, \dots, h_{1,1}({\cal M}_6)$). Therefore, there always remains at least one linear combination of the dilaton and K\"ahler moduli which does not couple to the (anomalous) $U(1)_a$'s. A full stabilization of the closed string moduli in a supersymmetric vacuum can therefore not be achieved by magnetic fluxes only. It may however be achieved by the introduction of three-form fluxes~\cite{AKM2}.

 \subsection{Non-supersymmetric vacua}

Let us now study the existence of consistent non-supersymmetric vacua $\langle d_a\rangle \neq 0$ induced by magnetized branes. Here, the cancellation of all R-R tadpoles does not anymore imply the cancellation of all tension contributions. On the contrary, a positive contribution $\delta T$ to the scalar potential arises from the sum over all tensions. Moreover,  the FI-parameter $\xi$ must be different than zero for the non-supersymmetric branes.
Altogether, the net contribution to the scalar potential on the disk is positive and the equation of motion for the dilaton cannot be satisfied. As it stands, non-supersymmetric magnetized branes do not lead to a consistent vacuum. 

Here, we propose two solutions to this problem. On the one hand, the closed string background may contain three-form fluxes.  A Gukov-Vafa-Witten superpotential is then generated~\cite{Giddings:2001yu,Gukov:1999ya}. In this case, assuming that minimal ${\cal N}=1$ supersymmetry is preserved in the bulk,  the dilaton is determined in terms of the NS-NS and R-R flux quanta~\cite{Kachru:2002he}. In a second step, supersymmetry is broken on some branes\footnote{This can also be applied to models where the moduli are stabilized in a Minkowski vacuum \cite{Becker:2007dn}.}. This generates a disk-level contribution to the scalar potential.  The dilaton's equation of motion is nevertheless satisfied perturbatively if a small hierarchy exists between the brane (disk) and the flux contributions (tree). 

This however forms  a consistent scenario under some strong constraints.  First, the Freed-Witten anomaly drastically restricts the configuration of branes~\cite{Freed:1999vc,Maldacena:2001xj}. For instance, $D9$ branes are forbidden. Second, three-form fluxes preserve the same supercharge as the $O3$ planes, but do not form a supersymmetric configuration with $O9$ planes. Consistent scenari must then involve magnetized $D7$ branes in an orientifold compactification with $O3$ planes~\cite{AKM2}. Third, one may wonder if the D-term breaking considered here is consistent with ${\cal N}=1$ supergravity constraints~\cite{Choi:2005ge}. Contrary to the standard case of global supersymmetry, local supersymmetry forbids pure D-term supersymmetry breaking. An uplift from an original AdS supersymmetric vacuum by pure D-term is then impossible. Combined effects of F- and D-term breaking must then be considered~\cite{Cremades:2007ig}. The scenario presented here is different. Indeed, the stabilization of K\"ahler moduli is achieved without any non-perturbative effects. Imaginary self-dual (ISD) three-form fluxes and constant internal magnetic fields can stabilize the vacuum in Minkowski space. Unlike in Ref.~\cite{Burgess:2003ic}, the value of the superpotential before uplifting  vanishes, $\langle W\rangle = 0$, and the argument of Ref.~\cite{Choi:2005ge} does not apply anymore. A dS uplifting by pure D-terms can then be achieved. 

The above constructions must  satisfy a last constraint: non-supersymmetric branes usually contain tachyonic modes in their open string spectra that signal instabilities. For instance, magnetized branes that do not preserve the same supersymmetry may  contain tachyons in their twisted sectors \cite{Bachas:1995ik}. These instabilities must be absent of consistent vacua. Examples of such models were presented in Ref.~\cite{Maillard:2007pq}.  It was shown that in particular regions of the closed string moduli space, the squared-masses of all twisted open string scalars are non-negative\footnote{A similar analysis including non-perturbative effects has been done in Ref.~\cite{GarciadelMoral:2005js}.}. 

The second solution involves non-critical strings. 
In addition to the disk contribution, the scalar potential acquires a contribution on the sphere arising  from the central charge deficit $\delta c$ of the conformal field theory (CFT).  Non-critical strings allow the presence of AdS vacua where the dilaton field is fixed at a perturbative regime\footnote{A similar phenomenon has been studied in non-critical type 0B string by Ref.~\cite{Israel:2007nj}.}. 

Let us now be more precise and derive explicitly the above statements. We start by considering the disk contribution to the scalar potential for a consistent set of $K+1$ magnetized branes.
 Let us assume that the first $K$ stacks stabilize the metric moduli by supersymmetric conditions. The only remaining massless scalar field from the closed string sector is the dilaton and the corresponding R-R axion. There is also a single massless $U(1)$ vector boson from the last magnetized brane with flux ${\cal F}_{K+1}\equiv {\cal F}$. 
 
The scalar potential (\ref{sec:DB;V}) is then the sum of three different terms.  Upon the R-R tadpole cancellation conditions, the first term  is  the tension deficit $\delta T$ of the last non-supersymmetric  brane, the second term comes from the DBI-action (\ref{BIdterms}) of the remaining massless $U(1)$, and the last contribution arises from its FI-term proportional to the parameter $\bar{\xi}$. Together, they can be written as
\beq
{\cal V}(\varphi, d) = T_3\, e^{-\varphi}\frac{2\pi }{ g^{2}_{K+1}}\Big[ \delta T -   \left(1- \sqrt{1- (2\pi\alpha' d)^2}\right)  + \bar{\xi} (2\pi\alpha' d)\Big] \equiv T_3\, e^{-\varphi}\, \delta \bar{T} \, , 
\label{pot1d}
\eeq 
where 
\beq
\delta T = 1- \sin x \, , \qquad \quad \bar{\xi} = \cos x \,  , \qquad \sin x =
\frac{A }{ \sqrt{A^2 + B^2}} 
\label{Txi}
\eeq
and 
\beq
A = \frac{1}{(4\pi^2\alpha')^3}  \int \left(J^3 - {\cal F}^2 J \right) \quad  {\rm and } \quad 
B= \frac{1}{(4\pi^2\alpha')^3} \int \left({\cal F}^3 - J^2 {\cal F} \right)\, .
\eeq

After elimination of the auxiliary field $d$, one easily realizes that  the potential is positive semi-definite
\beq
\delta \bar{T} = \frac{2\pi}{g^{2}_{K+1}}  \left( 1-\sin x - 1 + \sqrt{1+\cos^2 x}\right) =\sqrt{A^2 + B^2} \left( \sqrt{1+\bar{\xi}^2} - \sqrt{1- \bar{\xi}^2}\right) \; >0 \, .
\label{barT}
\eeq 
The only solution to the dilaton's equation of motion corresponds to the supersymmetric configuration where $\langle d\rangle= \bar{\xi}=\delta T = 0$ as in the quadratic approximation. This possibility is however excluded in our example with vanishing VEV's for open string charged states. Note that even in the case where the four-dimensional background is not Minkowski, but has a constant curvature, the equation of motion for the dilaton and the Einstein equations for the metric are incompatible.  It is therefore not possible to obtain consistent non-supersymmetric configurations of magnetized branes, neither in the Minkowski nor in the (A)dS space. 

In the presence of three-form closed string fluxes, the dilaton can be stabilized in a supersymmetric way by minimizing the tree-level potential induced by F-terms. In this case, the disk contribution (\ref{pot1d}) arising from the FI D-term is a positive constant and the Einstein equations for the metric are satisfied in a dS space-time with positive curvature, setup by $\delta{\bar T}$:
\beq
R=2\frac{T_3e^{3\varphi_0}\delta{\bar T}}{M_P^2}= \frac{e^{\varphi_0}\delta{\bar T}} {(2\pi)^2v_6}M_s^2\; >0\, ;\qquad v_6\equiv \frac{V_6}{(4\pi^2\alpha')^3}\, ,
\label{RdS}
\eeq
where 
$\varphi_0$ is the VEV of the dilaton and we used the expression (\ref{MP}) for the Planck mass. Supersymmetry is broken by a D-term of the $(K+1)$-th brane, given by:
\beq
\langle d\rangle = \frac{1}{2\pi\alpha'} \frac{\bar\xi}{\sqrt{1+{\bar\xi}^2}}\, ,
\label{dvalue}
\eeq
which in principle can be made small compared to the string scale by tuning the fluxes. This mechanism provides a solution to the so-called vacuum uplifting problem in the KKLT context~\cite{KKLT}.

In the absence of three-form fluxes, a different vacuum can be found by going off-criticality. Indeed,
for non-critical strings, an additional contribution to the scalar potential appears at the sphere-level, proportional to the central charge deficit $\delta c$. Together with the disk contribution, the scalar potential acquires the form
 \begin{eqnarray}
 {\cal V}_{nc}(\varphi) &=& e^{-2\varphi}\, v_6\, \delta c +  e^{-\varphi}\, T_3\, \delta \bar{T}
 \nonumber\\
 &=& e^{-2\varphi_4}\, \delta c + e^{-\varphi_4}\, v_6^{-1/2}\, T_3\, \delta \bar{T}
 \label{potnc}
 \end{eqnarray}
 in the string frame $(M_s=1)$, or equivalently,
 \beq
 {\cal V}_{nc}(\varphi) = e^{2\varphi_4} \delta c +  
 \frac{e^{3\varphi_4}}{(2\pi)^3v_6^{1/2}}\, \delta \bar{T}
 \label{potncE}
\eeq
in the Einstein frame $(M_P=1)$. Here $\varphi_4$ is the four-dimensional dilaton 
related to the ten-dimensional dilaton $\varphi$ by $e^{-2\varphi_4}=e^{-2\varphi}v_6$, with 
$v_6$ the six-dimensional volume given in eq.~(\ref{RdS}).
The potential (\ref{potnc}) has a minimum for a positive string coupling $g_s=e^{\varphi_0}$, with $\varphi_0=\langle\varphi\rangle$, only if $\delta c$ is negative: $\delta c<0$. In this case, the value of the potential at the minimum is also negative. We have
\beq
{\cal V}(\varphi = \varphi_0) = \frac{4}{27} \frac{ {\delta c}^3}{{\delta \bar{T}}^2}(2\pi)^6 v_6M_P^4 \; <0 \quad \quad {\rm and } \quad \quad e^{\varphi_0} = -\frac{2\delta c } 
{3\delta \bar{T} }(2\pi)^3v_6 \; >0 
\, .
\label{dil}
\eeq 

The scalar potential has therefore a non-supersymmetric minimum with a D-term supersymmetry breaking given by eq.~(\ref{dvalue}) and a negative vacuum energy. This solution corresponds to an AdS vacuum whose curvature may be given in terms of the fluxes and $\delta c$ as~\cite{ABEN}:
 \beq
 R =   
 \frac{8}{27}\frac{ {\delta c}^3}{{\delta \bar{T}}^2}(2\pi)^6 v_6M_P^2= \frac{2}{3\pi}\delta c M_s^2 \, ,  \qquad g_s= e^{\varphi_0} = - \frac{2\delta c}{3\delta \bar{T} }(2\pi)^3v_6 \, ,
 \eeq 
where $\delta \bar{T}$ is given in eq.~(\ref{barT}), in terms of volume moduli and fluxes of the non-supersymmetric brane. One sees that the string coupling $g_s$ can be made arbitrarily small by an adequate choice of CFT with small negative central charge deficit $\delta c$. Similarly, for fluxes and values of the moduli at their minimum such that $\delta \bar{T}$ is large, the string coupling can be fixed at a perturbative regime. This may be achieved together with a perturbatively small gauge coupling $g^{2}_{K+1}e^{\varphi_0} = e^{\varphi_0}/\sqrt{A^2+B^2}$ for the last non-supersymmetric brane. The AdS curvature can also be tuned in the same way.  In the perturbative regime where $g_s \ll 1$, this is also small provided the one-loop potential contribution $\delta \bar{T}$ is not too large.

Note that $\delta c$ can be done infinitesimally small only for negative values that are required here. The reason is that unitary CFT's can have accumulation points for their central charge only from below. It is then expected that $\delta c$ is quantized but can take infinitesimally small values. One simple example is provided by replacing a free compactified coordinate with a CFT from the minimal series. It would be of course very interesting to study explicitly the (closed and open) string quantization in this setup.

As shown in Section~\ref{secN=2sugra}, a non-critical dilaton potential, proportional to $\delta c$ in eq. (\ref{potnc}), is described by a gauging of the effective ${\cal N}=2$ supergravity of the closed string sector. Indeed, by considering the single dilaton hypermultiplet (in a vacuum where all other closed string moduli are fixed) and gauging the isometry associated to the shift of the NS-NS four-dimensional antisymmetric tensor using the graviphoton, one obtains the scalar potential (\ref{dilatonpot}). Identifying $V=e^{-\varphi_4}$ and the coupling constant of the gauging with $\delta c$, $g_*^2=-\delta c$, one obtains that the dilaton is stabilized at 
\beq
g_s = \frac{2}{3} \, g_*^2 \,  {\delta \bar{T}}^{-1}(2\pi)^3v_6
= \sqrt\frac{2}{3} (2\pi)^{3/2}v_6^{1/2} \, g_* \, g \left( \sqrt{1+\bar{\xi}^2} - \sqrt{1-\bar{\xi}^2} \right)^{-1/2} \; ,
\label{ads:gaugedstab}
\eeq
where $g$ is the physical gauge coupling of the non-supersymmetric brane.\footnote{Note that $g$ differs in general from the gauge couplings of the Standard Model which may arise on different set of branes.}
The $\bar{\xi}$-dependent term in eq.~(\ref{ads:gaugedstab}) is bounded, since $\bar{\xi} \in [0,1]$. In the  limit of vanishing FI-parameter $\bar{\xi}\rightarrow 0$, supersymmetry is restored on the entire set of branes, the disk amplitude vanishes and we end up with the sphere contribution which leads to a runaway behaviour for the dilaton field. At finite values of $\bar{\xi}$ however, supersymmetry is broken. A perturbative regime can then be found when  the  gauge coupling $g$ is small, or equivalently from eq.~(\ref{susy2}), when the volume of the internal manifold is stabilized at a relatively large value. 

It is important to notice that the validity of the approximation which allows us to fix the dilaton VEV by the method presented above relies on a perturbative expansion around the critical dimension for $\delta c$ small, together with the string loop  expansion for $g_s$ small, in a way that $g_s$ and $\delta c$ are the same order. Higher-order corrections can then be consistently neglected in the solution (\ref{dil}), under the usual assumption that there are no large numerical coefficients involved. 

The supersymmetry breaking solutions described above, with all closed string moduli stabilized (even in toroidal type I string compactifications), may be used for building simple models with interesting phenomenology. Indeed, Ref.~\cite{Antoniadis:2007jq} provides an example of a supersymmetric $SU(5)$ grand unified gauge group with three generations of quarks and leptons. As was pointed out in this work, the set of branes with VEV's for charged scalars needed to restore supersymmetry may be replaced by a brane sector where supersymmetry is broken by D-terms, while the dilaton is stabilized in a dS or AdS vacuum. This sector can be used as a source of supersymmetry breaking, mediated to the observable world by gauge interactions~\cite{Antoniadis:2006uj}. An obvious advantage of this framework is its calculability at the string level.

\section*{Acknowledgements}
We would like to thank Pablo Camara, Gianguido Dall'Agata, Emilian Dudas and Sergio Ferrara for very useful discussions.
This work was supported in part by the European Commission under the RTN contracts 
MRTN-CT-2004-503369 and MRTN-CT-2004-005104, in part by a European Union Excellence Grant
MEXT-CT-2003-509661, in part by the INTAS contract 03-51-6346, and in part by the CNRS contracts PICS \#~2530,  3059 and 3747.
The work of JPD was supported by the Swiss National Science Foundation.

\newpage
\appendix

\section{Conventions for \boldmath{${\cal N}=1$} superspace}\label{A1}

The ${\cal N}=1$ supersymmetry variation of a superfield $V$ is
$\delta V = ( \epsilon Q + \ov\epsilon\ov Q )V$, with supercharges 
\beq
\label{conv2}
Q_\alpha = \frac{\partial}{\partial \theta^\alpha} + i(\sigma^\mu\ov\theta)_\alpha
\, \partial_\mu \, , 
\qquad\qquad
\ov Q_\dalpha = -\frac{\partial}{\partial \ov\theta^\dalpha} 
- i(\theta\sigma^\mu)_\dalpha \, \partial_\mu \, ,
\eeq
where $\theta, \ov\theta$ are the Weyl spinor coordinates of the ${\cal N}=1$ superspace and 
$\sigma^\mu=(1\!\! 1, \sigma^i)$ with $1\!\! 1$ the identity and $\sigma^i$ the three Pauli matrices.
Since
\beq
\label{conv3}
\{ÊQ_\alpha, \ov Q_\dalpha\} =  -2i (\sigma^\mu)_{\alpha\dalpha} \, \partial_\mu \, ,
\eeq
the supersymmetry algebra is
\beq
\label{conv4}
[ \delta_1 , \delta_2 ] V = -2i \, ( \epsilon_1\sigma^\mu\ov\epsilon_2 
- \epsilon_2\sigma^\mu\ov\epsilon_1 ) \, \partial_\mu V.
\eeq

The covariant derivatives
\beq
\label{conv5}
D_\alpha = \frac{\partial}{\partial \theta^\alpha} - i(\sigma^\mu\ov\theta)_\alpha
\, \partial_\mu \, , 
\qquad\qquad
\ov D_\dalpha = \frac{\partial}{\partial \ov\theta^\dalpha} 
- i(\theta\sigma^\mu)_\dalpha \, \partial_\mu 
\eeq
anticommute with the supercharges and verify
\beq
\label{conv6}
\{ÊD_\alpha, \ov D_\dalpha\} =  -2i (\sigma^\mu)_{\alpha\dalpha} \, \partial_\mu 
\eeq
as well. The identities
\beq
DD\,\theta\theta = \ov{DD}\,\ov{\theta\theta} = -4, 
\qquad\qquad
\Dint = -\frac{1}{4}\Fint \ov{DD} = -\frac{1}{4}\Fbarint DD,
\eeq
valid under a space-time integral $\int d^4x$, are commonly used.

The super-Maxwell Lagrangian is 
\beq
\label{conv7}
{\cal L}_{Max.} = \frac{1}{4} \Fint WW + \frac{1}{4} \Fbarint \ov{WW} \, ,
\eeq
with
\beq
\label{conv8}
W_\alpha = - \frac{1}{4}\ov{DD}\, D_\alpha {\cal A},
\qquad\qquad
\ov W_\dalpha = - \frac{1}{4} {DD}\, \ov D_\dalpha {\cal A},
\eeq
and ${\cal A}$ is real. In this convention, $\ov W_\dalpha$ is {\it minus} the conjugate
of $W_\alpha$:
\beq
\label{conv9}
W_\alpha = -i\lambda_\alpha + \ldots
\qquad\qquad
\ov W_\dalpha = -i\ov\lambda_\dalpha + \ldots
\eeq
where $\lambda$ is the gaugino spinor. Then
\beq
\label{conv10}
\begin{array}{rcl}
WW &=& -\lambda\lambda + \ldots + \theta\theta[d^2 - \frac{1}{2}F_{\mu\nu} F^{\mu\nu}
- \frac{i}{2}F_{\mu\nu} \tilde F^{\mu\nu} 
+ 2i \lambda\sigma^\mu\partial_\mu\ov\lambda ] \, ,
\crbig
\ov{WW} &=& -\ov{\lambda\lambda} + \ldots + \ov{\theta\theta}[d^2 
- \frac{1}{2}F_{\mu\nu} F^{\mu\nu} + \frac{i}{2}F_{\mu\nu} \tilde F^{\mu\nu}
- 2i \partial_\mu\lambda\sigma^\mu\ov\lambda ]
\end{array}
\eeq
and
\beq
\label{conv11}
{\cal L}_{Max.} = \frac{1}{2} d^2 - \frac{1}{4}F_{\mu\nu} F^{\mu\nu}
+ \frac{i}{2}\lambda\sigma^\mu\partial_\mu\ov\lambda
- \frac{i}{2}\partial_\mu\lambda\sigma^\mu\ov\lambda \, .
\eeq

For a chiral superfield $\phi(y,\theta) = z(y) +\sqrt2 \, \theta\psi(y) - \theta\theta f(y)$, 
the ${\cal N}=1$ supersymmetry variations are
\beq
\label{conv12}
\begin{array}{rcl}
\delta z &=& \sqrt 2 \, \epsilon\psi \, , 
\crbig
\delta \psi_\alpha &=& -\sqrt2 \, [ f \epsilon_\alpha  
+ i(\sigma^\mu\ov\epsilon)_\alpha \partial_\mu z] \, ,
\crbig
\delta f &=& -\sqrt2 \, i \, \partial_\mu\psi\sigma^\mu\ov\epsilon.
\end{array}
\eeq

\section{Useful identities}\label{A2}

With $1=\epsilon^{12}=\epsilon^{\dot1\dot2}= -\epsilon_{12}=-\epsilon_{\dot1\dot2}$,
$$
\begin{array}{ll}
D_\alpha D_\beta = \frac{1}{2}\epsilon_{\alpha\beta} DD ,
\qquad\qquad
&\ov D_\dalpha \ov D_\dbeta = -\frac{1}{2}\epsilon_{\dalpha\dbeta}\ov{DD} ,
\crbig
[ D_\alpha , \ov{DD} ] = -4i(\sigma^\mu\ov D)_\alpha \partial_\mu ,
\qquad\qquad
&[ \ov D_\dalpha , DD ] = +4i(D\sigma^\mu)_\dalpha \partial_\mu ,
\crbig
DD\, W_\alpha = 4i(\sigma^\mu\partial_\mu\ov W)_\alpha ,
\qquad\qquad
&\ov{DD}\, \ov W_\dalpha = - 4i(\partial_\mu W\sigma^\mu)_\dalpha    ,
\crbig
[ DD , \ov{DD} ] = - 8i \, (D\sigma^\mu\ov D) \, \partial_\mu + 16 \, \Box
\,\, = & 8i \, (\ov D\ov\sigma^\mu D) \, \partial_\mu - 16 \, \Box  .
\end{array}
$$
Since
$$
\begin{array}{rcl}
D_\alpha \ov D_\dalpha D_\beta - D_\beta \ov D_\dalpha D_\alpha 
&=& -\frac{1}{2}\epsilon_{\alpha\beta} (\ov D_\dalpha DD + DD \ov D_\dalpha) ,
\crbig
\ov D_\dalpha D_\alpha \ov D_\dbeta - \ov D_\dbeta D_\alpha \ov D_\dalpha 
&=& \frac{1}{2}\epsilon_{\dalpha\dbeta} (D_\alpha \ov{DD} + \ov{DD} D_\alpha),
\end{array}
$$
we also have
$$
\begin{array}{rcl}
D^\alpha \ov D_\dalpha D_\alpha &=& 
-\frac{1}{2} (\ov D_\dalpha DD + DD \ov D_\dalpha) ,
\crbig
\ov D_\dalpha D_\alpha \ov D^\dalpha  
&=& -\frac{1}{2}(D_\alpha \ov{DD} + \ov{DD} D_\alpha).
\end{array}
$$
On a chiral superfield,
$$
\begin{array}{rcl}
D^\alpha \ov D_\dalpha D_\alpha \, \phi
&=& - \frac{1}{2}\ov D_\dalpha DD\phi ,
\crbig
\ov D_\dalpha D_\alpha \ov D^\dalpha \, \ov\phi
&=& -\frac{1}{2} D_\alpha \ov{DD} \ov\phi.
\end{array}
$$
For any chiral spinor superfield $\psi$,
$$
(\psi\sigma^\mu\ov\eta) \, \partial_\mu(\psi\psi) =
- (\partial_\mu\psi\sigma^\mu\ov\eta) \, \psi\psi.
$$

It is useful to notice that
$$
[ \ov{DD} , \eta D + \ov\eta\ov D ] =
\eta^\alpha [ \ov{DD} , D_\alpha ] = 4i \, (\eta\sigma^\mu \ov D) \, \partial_\mu.
$$
Hence, applying $\ov{DD}$ on the variation $\delta^*$  of a superfield
is not the same as the variation $\delta^*$ of $\ov{DD}$ applied on the same superfield,
except if the superfield is chiral or under a space-time integral $\int d^4x$.

\section{Solution of the constraint (\ref{*constraint})}\label{A3}

The nonlinear constraint (\ref{*constraint}) can be rewritten as:
\beq
\label{rewrite}
\kappa^2 X = WW - \frac{1}{2}\ov{DD} \frac{WW \, \ov{WW} }
{(\kappa^2+ \frac{1}{2} \ov{DD}\, \ov X)(\kappa^2 + \frac{1}{2}DDX)}.
\eeq
To find $X$, we need to find an expression for $DDX$ in the denominator.
In general 
$$
DDX = DD\frac{WW} {\kappa^2 + \frac{1}{2}\ov {DD}Ê\ov X} ,
$$ 
but we know that in the denominator of expression (\ref{rewrite}), the derivatives
must act on $WW$: any other choice would lead to a factor $W_\alpha$ or 
$\ov W_\dalpha$ in the expansion of the denominator and then to a 
vanishing contribution since $W_\alpha W_\beta W_\gamma=0$. 
It is then sufficient to solve the simple equation 
\beq
DDX = \frac{1}{\kappa^2+\frac{1}{2} \ov{DD}\, \ov X} \, DDWW.
\eeq
The solution is
\beq
\ov{DD}\ov X = -\kappa^2\left[ 1+B - \sqrt{1 + 2A + B^2} \right],
\eeq
with 
$$
A = \frac{1}{2\kappa^4} ( DD\, W^2  + \ov{DD} \,\ov W^2) = A^*, \qquad\qquad
B = \frac{1}{2\kappa^4} ( DD\, W^2  - \ov{DD} \,\ov W^2) = -B^*.
$$
This solution can then be inserted in the denominator of eq.~(\ref{rewrite}),
to obtain the final expression (\ref{Xis}).

In order to derive the component expression of eq.~(\ref{BI}) of the Lagrangian (\ref{Xis}), 
one uses the identities:
$$
\begin{array}{c}
(F_{\mu\nu}\tilde F^{\mu\nu})^2 
= \frac{1}{4} (\epsilon_{\mu\nu\rho\sigma}F^{\mu\nu}F^{\rho\sigma})^2
= 4F_{\mu\nu}F^{\nu\rho}F_{\rho\sigma}F^{\sigma\mu}
- 2 (F_{\mu\nu}F^{\mu\nu})^2 ,
\crbig
-{\rm det}\left( \eta_{\mu\nu} + A F_{\mu\nu} \right)
= 1 + \frac{A^2}{2} F_{\mu\nu}F^{\mu\nu}
- \frac{A^4}{16} (F_{\mu\nu}\tilde F^{\mu\nu})^2 .
\end{array}
$$

%%%%%%%%%%%%%

\end{document}